\documentclass[11pt,a4paper]{article}

\usepackage{jheppub}
\usepackage{subfig}
\usepackage{amsmath}
\usepackage{graphicx}

\def\eps{\epsilon}
\def\hh{\lambda}
\def\la{\langle}
\def\ra{\rangle}
\def\tP{\tilde{P}}
\def\rvec{\eta}
\def\smallpar{\delta}

\def\trF{{\rm tr}}
\def\tf{\tilde{f}}

\def\spA#1#2{\la #1 #2 \ra}
\def\spB#1#2{[ #1 #2 ]}

\def\spAB#1#2#3{\la #1 | #2 | #3 ]}

\def\zspA#1{\la z_{#1} \ra}
\def\zspB#1{[ z_{#1} ]}
\def\spr{\alpha}
\def\sptr{\beta}
\def\spX{\gamma}

\def\wspA#1{\la \omega_{#1} \ra}
\def\wspB#1{[ \omega_{#1} ]}

\def\to{\rightarrow}
\def\mcoll#1{\overset{1||\dots||#1}{\to}}

\def\Sp{{\rm Sp}}
\def\cSp{\mathcal{S}p}

\def\Fnmhv#1{{\rm F}^{\textsc{\tiny NMHV}}_#1}

\def\Fonem{{\rm F}^{1m}_{\rm box}}

\def\qb{{\bar{q}}}
\def\cN{\mathcal{N}}

\definecolor{mygreen}{rgb}{0,0.7,0}

\preprint{}

\title{One-loop triple collinear splitting amplitudes in QCD}

\author[a]{Simon Badger,}
\author[a]{Francesco Buciuni,}
\author[a]{Tiziano Peraro\,}
\affiliation[a]{
Higgs Centre for Theoretical Physics, %
School of Physics and Astronomy, %
The University of Edinburgh, %
Edinburgh EH9 3JZ, Scotland, UK
}
\emailAdd{Simon.Badger@ed.ac.uk, Tiziano.Peraro@ed.ac.uk, Francesco.Buciuni@ed.ac.uk}

\abstract{
We study the factorisation properties of one-loop scattering amplitudes
in the triple collinear limit and extract the universal splitting amplitudes
for processes initiated by a gluon.  The splitting amplitudes are derived from the analytic Higgs plus four partons amplitudes.  We present compact results for
primitive helicity splitting amplitudes making use of super-symmetric decompositions.
The universality of the collinear factorisation is checked numerically against the full colour
six parton squared matrix elements.
}

\preprint{Edinburgh 2015/12}

\begin{document}
\maketitle

\section{Introduction}

A full understanding of the infrared structure of QCD matrix elements is an unavoidable
step towards making precise predictions of Standard Model backgrounds at hadron colliders.
In order to make finite predictions for cross-sections we must cancel infrared singularities
between unresolved real radiation corrections and singularities in the virtual (loop)
corrections. The study of infrared properties of perturbative gauge theories have a broader scope
beyond this application since the universal behaviour provides a strong constraint on the structure
of scattering amplitudes.

The soft and collinear infrared limits at next-to-leading order (NLO) have been understood long ago
and general algorithms (e.g. Catani-Seymour \cite{Catani:1996vz} or
Frixone-Kunszt-Signer \cite{Frixione:1995ms}) for the computation of infrared finite cross-sections form the core
of the current generation of precision tools used to make theoretical predictions for the LHC
experiments.

In the last ten years or so a lot of effort has been put into generalising these techniques to
next-to-next-leading order (NNLO) and a variety of different techniques now exist with the ability
to make finite predictions for important LHC observables (e.g. references
\cite{GehrmannDeRidder:2005cm,Catani:2007vq,Czakon:2010td,Boughezal:2011jf,Czakon:2014oma,DelDuca:2015zqa}). All of these methods rely on knowledge
of the underlying factorisation properties of QCD amplitudes in the double unresolved limits at
tree-level \cite{Catani:1999ss} and single unresolved limits at one-loop
\cite{Bern:1994zx,Bern:1994fz,Kosower:1999rx,Bern:1998sc,Bern:1999ry}.

The first step at next-to-next-to-next-to-leading order (N${}^{3}$LO) has been taken recently through
the complete calculation of fully inclusive Higgs production at hadron colliders up to $\mathcal{O}(\alpha_s^5)$ in the large
top quark mass limit \cite{Anastasiou:2015vya}. This calculation has been performed in a number of
different stages building expansions around the soft limit
\cite{Anastasiou:2014vaa,Anastasiou:2014lda,Li:2014afw} and using the reverse unitarity method
to obtain each component of the triple-virtual \cite{Baikov:2009bg,Gehrmann:2010ue,Gehrmann:2010tu},
squared real-virtual \cite{Anastasiou:2013mca,Kilgore:2013gba}, double-virtual-real
\cite{Duhr:2013msa,Li:2013lsa,Duhr:2014nda,Dulat:2014mda},
double-real-virtual \cite{Li:2014bfa,Anastasiou:2015yha} and
triple-real radiation \cite{Anastasiou:2013srw} as an expansion in the dimensional regularisation parameter. The poles
of these separate contributions cancel analytically when summed together and combined with the counter-terms
for UV poles \cite{Tarasov:1980au,Larin:1993tp,vanRitbergen:1997va,Czakon:2004bu} and initial state
infrared singularities
\cite{Vogt:2004mw,Moch:2004pa,Anastasiou:2012kq,Hoschele:2012xc,Buehler:2013fha}.

Further steps are required to extend these techniques to fully differential observables in an
analogous way to the NLO and NNLO cases. Many of the infrared regions that must be accounted for in
such a procedure are now fully understood. The missing ingredients that remain are the one-loop
triple collinear splitting functions involving gluons.
Though the factorisation of the squared matrix elements are sufficient for the construction
of infrared finite cross sections, factorisation at the amplitude level \cite{Kosower:1999xi} can yield much more compact
expressions leading to a more efficient construction of the factorised squared matrix element,
especially when considering spin-correlations.

\begin{figure}[h]
  \begin{center}
    \includegraphics[width=8cm]{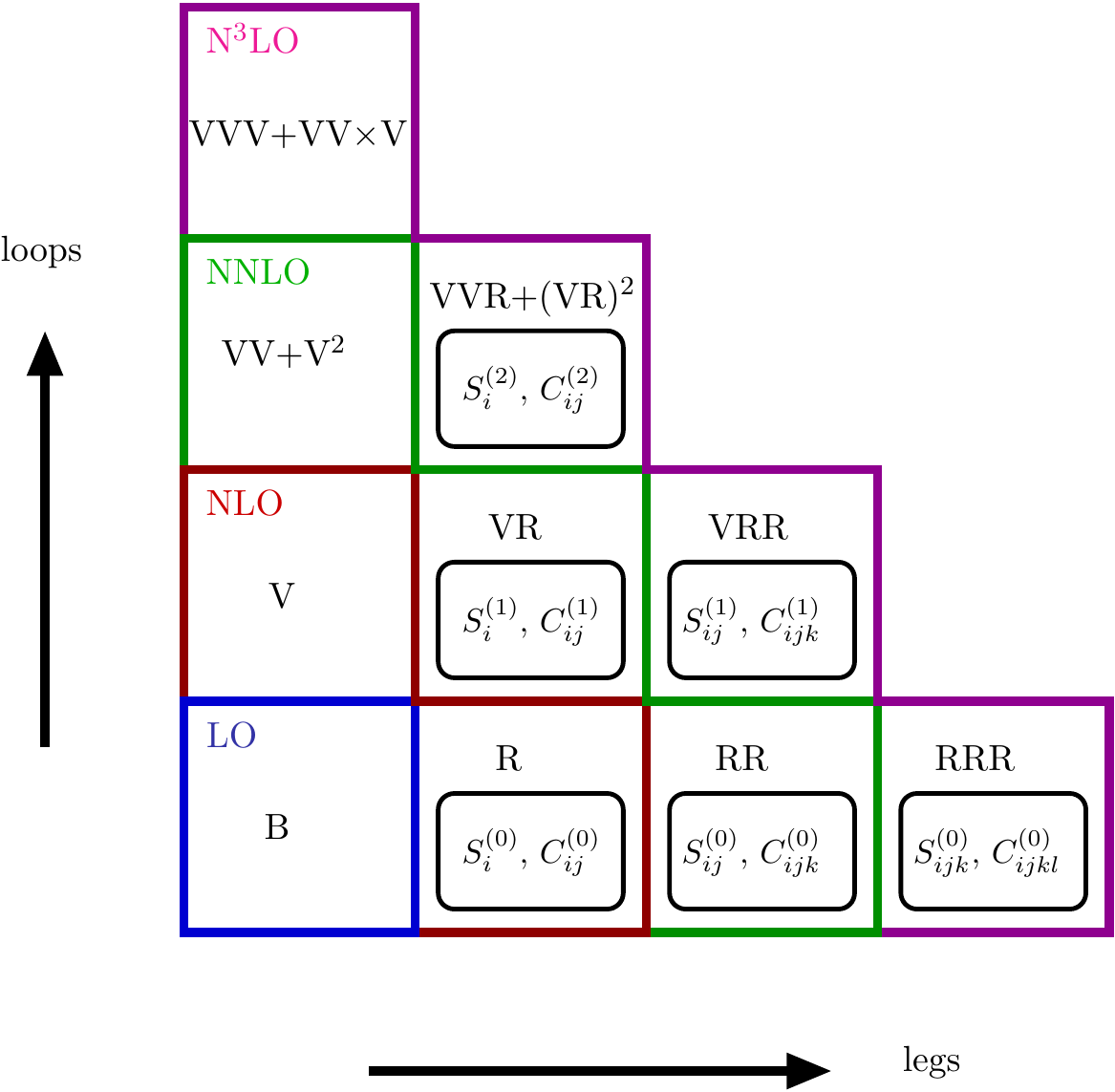}
  \end{center}
  \caption{The contributions to perturbative cross sections up to N${}^3$LO. This
  consists of virtual (V) corrections up to three loops and real radiation (R) corrections with
  up to 3 additional unresolved legs. In the real radiation contributions the primary infrared
  limits of soft (S) and collinear (C) should be removed from the matrix elements and re-combined
  with the virtual corrections to obtain an infrared finite result.
  }
  \label{fig:N3LO}
\end{figure}

Figure \ref{fig:N3LO} shows the real and virtual contributions to a cross section up to N${}^3$LO
and the primary singular limits which are either multiple soft, $S_{i_1\dots i_m}$, or multiple
collinear, $C_{i_1\dots i_m}$. The factorisation amplitudes have been computed in all
cases
\cite{Bern:2004cz,Badger:2004uk,Duhr:2013msa,Duhr:2014nda,Campbell:1997hg,Catani:1999ss,DelDuca:1999ha,Birthwright:2005ak,Birthwright:2005vi} except
for the triple-collinear and double-soft limits of the double-real-virtual. The triple collinear
limit at one-loop has been considered at the squared amplitude level for $q\to
q\bar{Q}Q$~\cite{Catani:2003vu} and
for the mixed QCD+QED cases of $q\to q\gamma\gamma$, $q\to qg\gamma$, $g\to \bar{q}q\gamma$,
$\gamma\to \bar{q}q\gamma$ and $\gamma\to \bar{q}qg$ \cite{Sborlini:2014mpa,Sborlini:2014kla}.

In this article we compute the one-loop gluon initiated splitting functions in QCD, $g\to ggg$ and
$g\to\qb q g$. Splitting amplitudes valid in four dimensions are extracted from the known analytic
amplitudes for $0\to Hgggg$
~\cite{Berger:2006sh,Badger:2006us,Badger:2007si,Glover:2008ffa,Badger:2009hw} and $0\to H \qb q gg$
~\cite{Berger:2006sh,Dixon:2009uk,Badger:2009vh} in the large top-mass limit.

The structure of the article is as follows. We first introduce the notation for the amplitudes and the
squared amplitudes together with their respective colour decompositions and collinear limits. In
section \ref{sec:limparam} we describe a parametrisation of the multi-collinear limit using
spinor-helicity variables which we will use to compute the splitting amplitudes. We then present the
$g\to ggg$ and $g\to \bar{q} q g$ splitting amplitudes and describe the symmetries and
super-symmetric decompositions used to obtain a compact representation. We then check the
universality of the new splitting amplitudes by taking a numerical limit of the $gg\to gggg$ and
$gg\to\bar{q}qgg$ in \textsc{NJet} before reaching our conclusions.

\section{Notation}

A general QCD amplitude can be decomposed into a basis of $SU(N_c)$
colour factors and ordered partial amplitudes which depend only on the
momenta and helicities of the external legs. For an $n$-point $L$-loop
amplitude this can be represented as,
\begin{equation}
  \mathcal{A}_n^{(L)}(\{a_i\},\{p_i^{\hh_i}\}) = \sum_c C_c(\{a_i\})
  A_{n;c}^{(L)}(\{p_i^{\hh_i}\})
  \label{eq:colourdecompositions}
\end{equation}
where $a_i, \hh_i$ and $p_i$ are colour indices (adjoint or
fundamental), helicity and momenta of the $i^{\rm th}$ leg.  Unless
explicitly indicated otherwise, we understand that the index $i$ runs
from $1$ to $n$, e.g.\
\begin{equation}
  \{p_i^{\hh_i}\} \equiv \{p_i^{\hh_i}\}_{i=1}^n = \{p_1^{\hh_1},\ldots,p_n^{\hh_n}\}.
\end{equation}
For cross-section computations we are required to square these
amplitudes and sum over the colour indices.  This sum can be
represented as,
\begin{align}
  \mathcal{M}_n^{(L,L')}(\{p_i^{\hh_i}\}) ={}& \sum_{a_i} \left(\mathcal{A}_n^{(L)}(\{a_i\},\{p_i^{\hh_i}\})\right)^\dagger \mathcal{A}_n^{(L')}(\{a_i\},\{p_i^{\hh_i}\})
  \nonumber\\
    ={}&
    \left(\vec{A}_n^{(L)}(\{p_i^{\hh_i}\})\right)^\dagger \cdot \mathcal{C}_n^{(L,L')} \cdot
    \vec{A}_n^{(L')}(\{p_i^{\hh_i}\}), \label{eq:Mnpartial}
\end{align}
where the matrix $\mathcal{C}_n^{(L,L')}$ is a function of $N_c$ defined
by
\begin{equation}
  \Big( \mathcal{C}_n^{(L,L')}\Big)_{c c'} = \sum_{a_i} \big(C_c(\{a_i\})\big)^\dagger  C_{c'}(\{a_i\}), \label{eq:colormatpartial}
\end{equation}
while $\vec{A}^{(L)}$ is a vector of partial amplitudes $A_{n;c}^{(L)}$
\begin{equation}
  \vec{A}^{(L)} = \{ A_{n;1}^{(L)}, A_{n;2}^{(L)}, \ldots \}.
\end{equation}

Partial amplitudes may in turn be written in terms of primitive
amplitudes $A_p^{[X]}$ which further decompose colour and flavour
structure due to the internal loops,
\begin{align}
  A_{n;c}^{(L)} = \sum_{p,X} R_{c,p,X}(N_c, N_f) A_{n,p}^{[L,X]}, \label{eq:partialvsprimitive}
\end{align}
where $X$ runs over the independent primitive topologies at $L$ loops
and $p$ runs over permutations of the $n$ external legs.
Eq.~\eqref{eq:Mnpartial} can thus be equivalently written as
  \begin{align}
  \mathcal{M}_n^{(L,L')} (\{p_i^{\hh_i}\}) =
    \left(\vec{A}_n^{[L]}(\{p_i^{\hh_i}\})\right)^\dagger \cdot \mathcal{C}_n^{[L,L']} \cdot
    \vec{A}_n^{[L']}(\{p_i^{\hh_i}\}) \label{eq:Mnprimitive}
\end{align}
where $\vec{A}_n^{[L]}$ is a vector of primitive amplitudes
$A_{n,p}^{[L,X]}$ and the matrix $\mathcal{C}_n^{[L,L']}$ can be
related to $\mathcal{C}_n^{(L,L')}$ defined in
Eq.~\eqref{eq:colormatpartial} by the change of basis in
Eq.~\eqref{eq:partialvsprimitive}.

\begin{figure}[t]
  \begin{align*}
    \parbox{3cm}{\includegraphics[width=3cm]{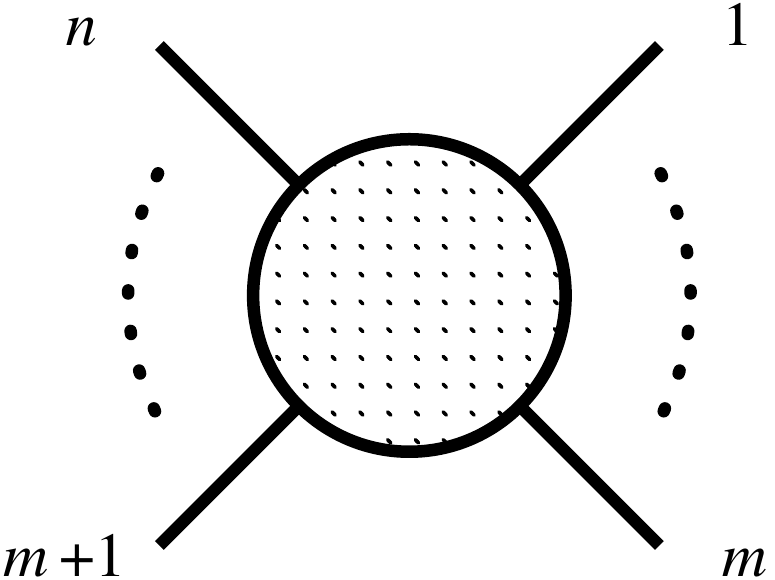}}
    \overset{1||\cdots||m}{\longrightarrow}
    \sum_{h=\pm}
    &\parbox{5cm}{\includegraphics[width=5cm]{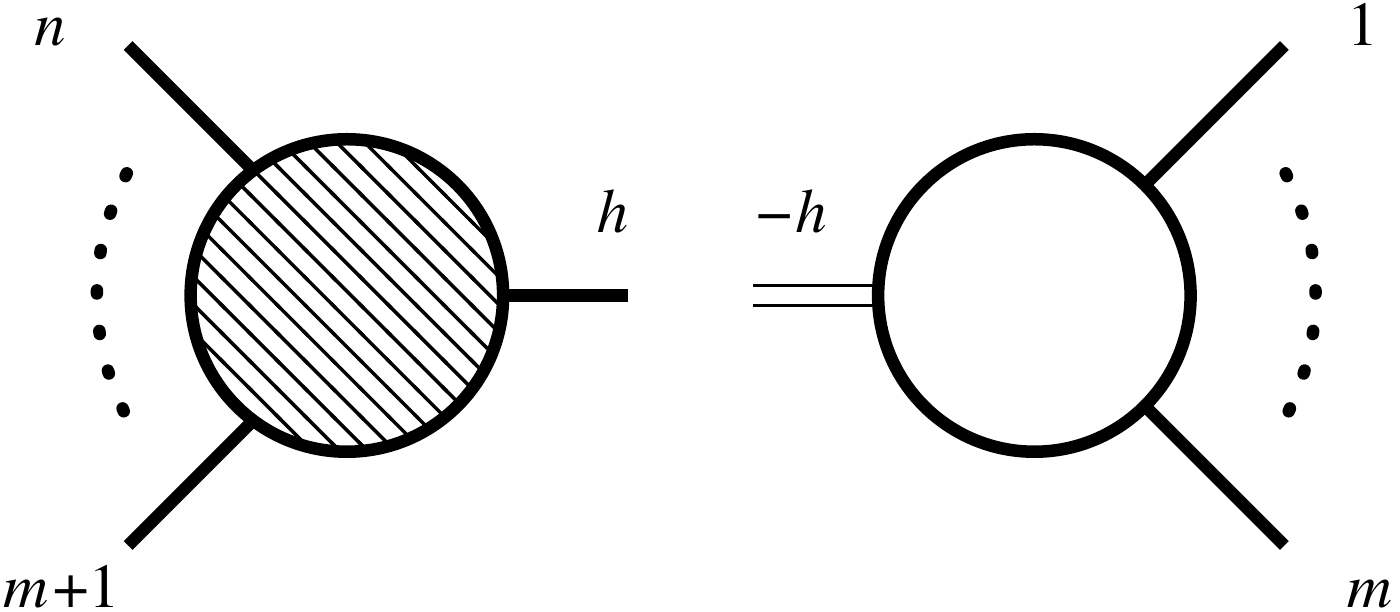}}
    \\
    \parbox{3cm}{\includegraphics[width=3cm]{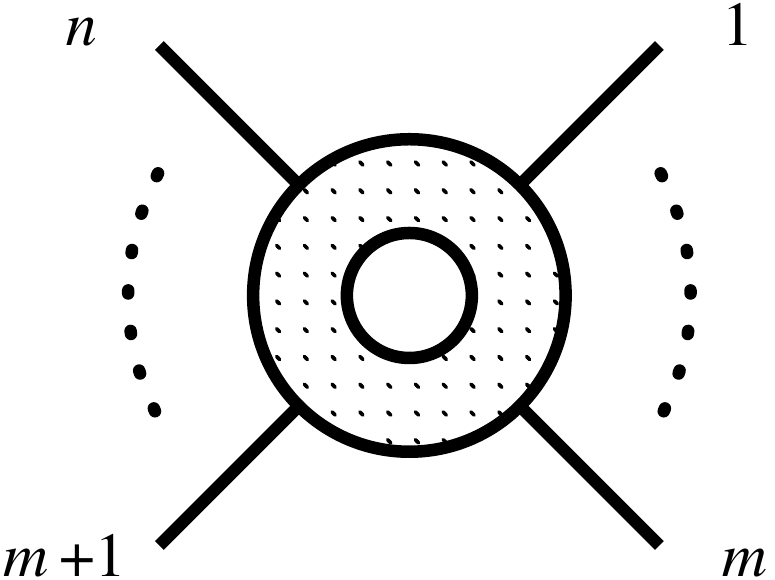}}
    \overset{1||\cdots||m}{\longrightarrow}
    \sum_{h=\pm}
    &\parbox{5cm}{\includegraphics[width=5cm]{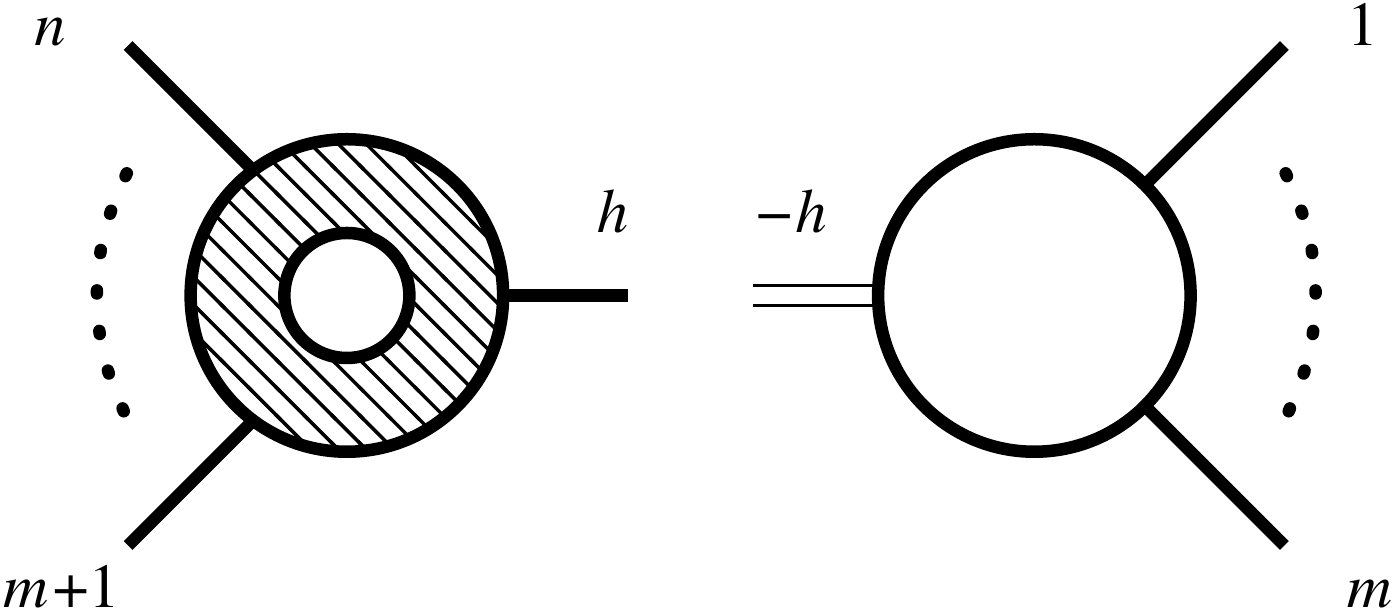}}\\
   +&\parbox{5cm}{\includegraphics[width=5cm]{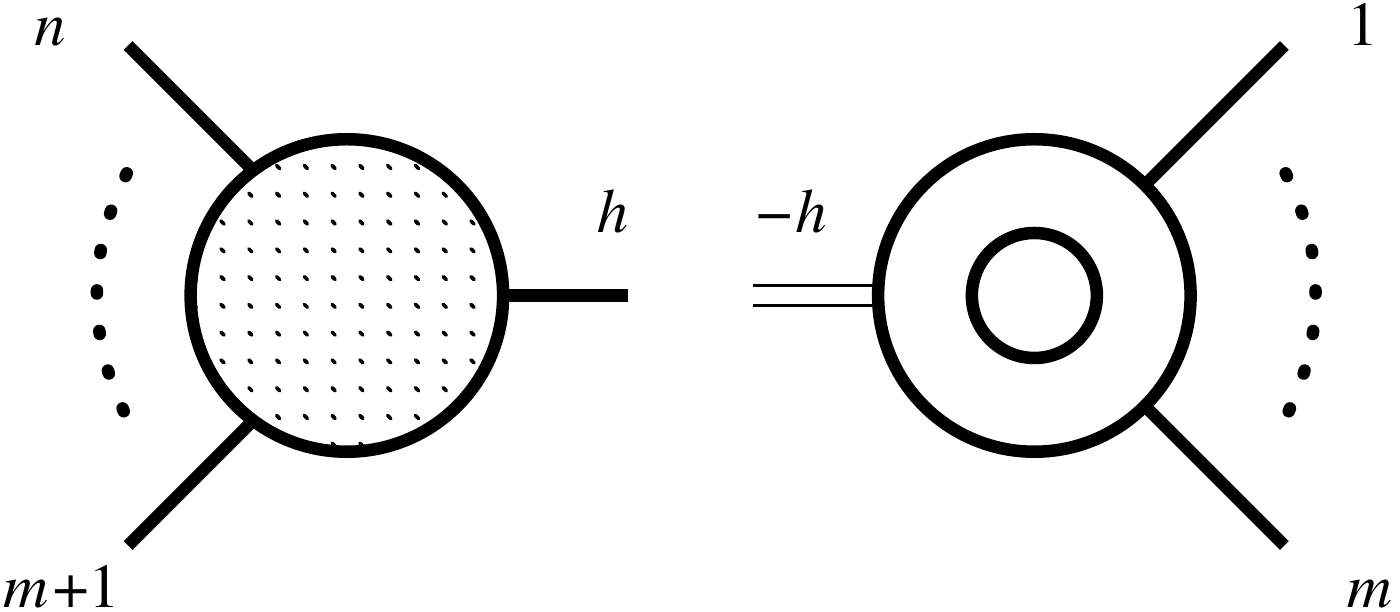}}
  \end{align*}
  \caption{Factorisation of tree and one-loop amplitudes in the multi-collinear limit.}
  \label{fig:clim}
\end{figure}

In the limit where $m$ of the external legs become simultaneously
collinear, the amplitudes factorise into a product of lower
multiplicity amplitudes and splitting amplitudes which contain all the
infrared divergences:
\begin{align}
  A_{n}^{(L)}(\{p_i^{\hh_i}\}) \mcoll{m}
  \sum_{k=0}^L \sum_{\hh_P}
  \Sp_m^{(L-k)}(-P^{-\hh_P}; \{p_i^{\hh_i}\}_{i=1}^m)
  A_{n-m+1}^{(k)}(P^{\hh_P},\{p_i^{\hh_i}\}_{i=m+1}^n)
  \label{eq:collfact}
\end{align}
where $A_{n}^{(L)}$ and $\Sp_n^{(L)}$ can either be primitive or
partial $n$-point amplitudes and splitting amplitudes respectively,
while and $P\equiv p_1+\cdots+p_m$. A schematic representation of this factorisation
is shown in Figure \ref{fig:clim}. The sum of internal helicity
states $\hh_P$ leads to spin correlations in the factorized
squared amplitude $\mathcal{M}^{(L,L')}$,
\begin{align}
  \mathcal{M}_n^{(L,L')}&(\{p_i^{\hh_i}\}) \mcoll{m}
  \sum_{k=0}^L \sum_{k'=0}^{L'} \sum_{\hh_P, \hh_P'} \nonumber\\&
  \mathcal{P}^{(L-k,L'-k')}_{m;-\hh_P,-\hh_P'}(-P; \{p_i^{\hh_i}\}_{i=1}^m)
  \, \mathcal{M}_{n-m+1;\hh_P,\hh_P'}^{(k,k')}(P, \{p_i^{\hh_i}\}_{i=m+1}^n)
  \label{eq:csumfactlimit}
\end{align}
where we can define
\begin{align}
  \mathcal{M}_{n;\hh_P,\hh_P'}^{(L,L')}(P,\{p_i^{\hh_i}\}) = {}&
  \left(\vec{A}_n^{(L)}(P^{\hh_P},\{p_i^{\hh_i}\})\right)^\dagger \cdot \mathcal{C}_n^{(L,L')} \cdot
  \vec{A}_n^{(L')}(P^{\hh_{P'}},\{p_i^{\hh_i}\}) \\
  \mathcal{P}_{n;\hh_P,\hh_P'}^{(L,L')}(P;\{p_i^{\hh_i}\}) = {}&
  \left(\vec{\Sp}_n^{(L)}(P^{\hh_P};\{p_i^{\hh_i}\})\right)^\dagger \cdot \mathcal{C}_{\Sp,n}^{(L,L')} \cdot
  \vec{\Sp}_n^{(L')}(P^{\hh_{P'}};\{p_i^{\hh_i}\}) \label{eq:csumfact}
\end{align}
in terms of partial amplitudes or equivalently
\begin{align}
  \mathcal{M}_{n;\hh_P,\hh_P'}^{(L,L')}(P,\{p_i^{\hh_i}\}) = {}&
  \left(\vec{A}_n^{[L]}(P^{\hh_P},\{p_i^{\hh_i}\})\right)^\dagger \cdot \mathcal{C}_n^{[L,L']} \cdot
  \vec{A}_n^{[L']}(P^{\hh_{P'}},\{p_i^{\hh_i}\}) \\
  \mathcal{P}_{n;\hh_P,\hh_P'}^{(L,L')}(P;\{p_i^{\hh_i}\}) = {}&
  \left(\vec{\Sp}_n^{[L]}(P^{\hh_P};\{p_i^{\hh_i}\})\right)^\dagger \cdot \mathcal{C}_{\Sp,n}^{[L,L']} \cdot
  \vec{\Sp}_n^{[L']}(P^{\hh_{P'}};\{p_i^{\hh_i}\}) \label{eq:csumfactprimitive}
\end{align}
in terms of primitive amplitudes.  In the colour matrix
$\mathcal{C}_{\Sp,n}^{[L,L']}$ we absorbed a prefactor which takes
into account colour conservation along the factorized parton, such that
\begin{equation}
  \mathcal{C}_{\Sp,n}^{[L,L']} =
  \left\{ \begin{array}{ll}
    \dfrac{1}{N_c^2-1} \mathcal{C}_n^{[L,L']} & \quad \text{for gluon-initiated $\Sp$} \\
    \dfrac{1}{N_c} \mathcal{C}_n^{[L,L']} & \quad \text{for quark-initiated $\Sp$} \\
  \end{array} \right. ,
\end{equation}
and similar for $\mathcal{C}_{\Sp,n}^{(L,L')}$.

For brevity, the results presented in this paper will often omit
the subscript indicating the number of partons involved in an
amplitude, since this can be deduced by its arguments, i.e.\
\begin{align}
  \Sp(-P^{\hh_P};p_1^{\hh_1},\ldots ,p_n^{\hh_n}) \equiv {}& \Sp_n(-P^{\hh_P};\{p_i^{\hh_i}\}) \nonumber \\
  A(p_1^{\hh_1},\ldots, p_n^{\hh_n}) \equiv {}& A_n(\{p_i^{\hh_i}\}).
\end{align}

\section{A spinor parametrisation of the multi-collinear
  limit \label{sec:limparam}}

We define the multiple collinear limit using a parametrisation of the
full kinematics in term of a parameter $\smallpar$, such that the
collinear limit in Eq.~\eqref{eq:collfact} is identified as the
leading term as $\smallpar \to 0$, i.e.\
\begin{align}
  & \underset{1||\cdots||m}{\lim} A_{n}^{(L)}(\{p_i^{\hh_i}\}) = \underset{\smallpar\to 0}{\lim} A_{n}^{(L)}(\{p_i^{\hh_i}(\smallpar)\}) \nonumber \\
  & \qquad ={} \sum_{k=0}^L \sum_{\hh_P}
  \Sp_m^{(L-k)}(-P^{-\hh_P}, \{p_i^{\hh_i}\}_{i=1}^m)
  A_{n-m+1}^{(k)}(P^{\hh_P},\{p_i^{\hh_i}\}_{i=m+1}^n) + \mathcal{O}\left(\frac{1}{\smallpar^{m-2}}\right).
  \label{eq:collfactlambda}
\end{align}
The parametrisation is defined by,
\begin{align}
  p_i^\mu(\smallpar) &= z_i \tP^\mu + \smallpar k^\mu_{T,i} - \smallpar^2 \frac{k_{T,i}^2}{2 (P\cdot \rvec) z_i}\rvec^\mu \label{eq:pidelta}
  & i&=1,\ldots,m \\
  p_i^\mu(\smallpar) &= K^\mu_{i}(\smallpar, \{p\}_{m+1,n},\rvec) & i&=m+1,\ldots,n
\end{align}
where $z_i = (p_i\cdot \rvec)/(P\cdot \rvec)$ are the momentum fractions of the unresolved partons,
$\rvec$ is an arbitrary light-like momentum and $\tP$ is the massless projection of $P = \sum_{i=1}^m p_i$,
\begin{equation}
  \tP^\mu = P^\mu - \frac{P^2}{2\,P\cdot \rvec} \rvec^\mu.
\end{equation}
The vectors $k^\mu_{T,i}$ are orthogonal to $P$, $\tP$ and $\rvec$
\begin{equation}
  k_{T,i} \cdot \tP = k_{T,i} \cdot P = k_{T,i} \cdot \rvec = 0.
\end{equation}
Momentum conservation implies that:
\begin{align}
  \sum_{i=1}^m z_i &= 1 \label{eq:sumzi}\\
  \sum_{i=1}^m k_{T,i}^\mu &= 0^\mu \\
  -\smallpar^2 \sum_{i=1}^m \frac{k_{T,i}^2}{z_i} &=  P^2.
\end{align}
The function $K^\mu_{i}$ is a generic map that keeps the factorized
momenta $m+1,\ldots,n$ on-shell as well as absorbing the recoil
$P^2/(2 P\cdot \rvec)\rvec^\mu$, and it satisfies $K^\mu_i\to p^\mu_i$
as $\smallpar \to 0$. The exact form is not important for our purpose
of explicitly taking the limit and various mappings have been considered
in the literature (for example in the Catani-Seymour subtraction \cite{Catani:1996vz} or Kosower's antenna
\cite{Kosower:1997zr}). When implementing the collinear phase-space numerically we employed
the Catani-Seymour map as described in Appendix \ref{app:collinearphasespace}.

Since we are working at the amplitude level, we would like to have a parametrisation
of the limit valid for the spinors of $p_i$ as well. This can be achieved using an
appropriate choice of the transverse vectors $k_{T,i}$,
\begin{equation}
  2\, \smallpar\, k^\mu_{T,i} =
    \zspA{i}\wspB{i} \spAB{\tP}{\gamma^\mu}{\rvec}
  + \zspB{i}\wspA{i} \spAB{\rvec}{\gamma^\mu}{\tP}.
  \label{eq:kTdef}
\end{equation}
In the above we use the notation
\begin{align}
  \zspA{i} = \frac{\spA{i}{\rvec}}{\spA{\tP}{\rvec}}, \qquad
  \zspB{i} = \frac{\spB{i}{\rvec}}{\spB{\tP}{\rvec}}, \qquad
  \wspA{i} = \frac{\spA{i}{\tP}}{\spA{\rvec}{\tP}}, \qquad
  \wspB{i} = \frac{\spB{i}{\tP}}{\spB{\rvec}{\tP}},
\end{align}
where the spinor variables $\zspA{i}$ and $\zspB{i}$ differ by a phase
from the usual parametrisation which uses $\sqrt{z_i}$.  It is worth
to notice that both $\wspA{i}$ and $\wspB{i}$ are
$\mathcal{O}(\smallpar)$ in the collinear limit.  The spinors
parametrisation then reads,
\begin{align}
  |i\ra &= \zspA{i}|\tP\ra + \wspA{i} |\rvec\ra &
  |i] &= \zspB{i}|\tP] + \wspB{i} |\rvec].
  \label{eq:collparamSP}
\end{align}
We find that this is a convenient way to take the limit at the
amplitude level since the spinor variables $\zspA{i}$ obey Schouten
identities:
\begin{equation}
  \sum_{ijk\,\mathrm{cyclic}} \zspA{i}\spA{j}{k} = 0,
  \label{eq:zschouten}
\end{equation}
as well as momentum conservation,
\begin{equation}
  \sum_i p_i^\mu - \tP^\mu - \frac{P^2}{2P\cdot \rvec} \rvec^\mu  = 0^\mu.
\end{equation}
For the triple collinear splitting amplitudes this means we have
the kinematics of a five-point function event though the colour space
is that of a four-point function.

\subsection{Example: the tree-level MHV multi-collinear splitting amplitude}

The result for the multi-collinear limit of the maximal-helicity-violating (MHV) amplitude
has been known for a long time. More recently the general helicity cases were also examined
through use of the MHV rules \cite{Birthwright:2005ak,Birthwright:2005vi}. This case
is incredibly straightforward and serves as a useful example of the general treatment introduced in the
previous section.

We start with the Parke-Taylor MHV amplitude with particles $1$ and $r>m$ having negative helicities
and all others positive helicity,
\begin{align}
  A^{(0)}_n(1^-,2^+,3^+,\dots,r^-,\dots,n^+) = \frac{\spA1r{}^4}{\prod_{i=1}^n \spA{i}{i+1}},
\end{align}
where the product in the denominator is considered modulo $n$. The limit is simply taken
by applying eq. \eqref{eq:collparamSP}
\begin{align}
  &A^{(0)}_n(1^-,2^+,3^+,\dots,r^-,\dots,n^+) =
  \nonumber\\&
  \frac{\left(\zspA{1}\spA{\tP_{1,m}}{r} + \wspA{1}\spA{\eta}{r} \right)^4}{
  \prod_{j=1}^{m-1} \spA{j}{j+1}
  \left( \zspA{1}\spA{n}{\tP_{1,m}} + \wspA{1}\spA{n}{\eta} \right)
  \left( \zspA{m}\spA{\tP_{1,m}}{m+1} + \wspA{m}\spA{\eta}{m+1}\right)
  \prod_{i=r}^{n-1} \spA{i}{i+1}}
  \nonumber\\&
  \xrightarrow[]{\smallpar\to0}
  \frac{\zspA1^3}{\zspA{m} \prod_{j=1}^{m-1} \spA{j}{j+1}} \,
  \frac{\spA{\tP_{1,m}}{r}^4}{\spA{n}{\tP_{1,m}}\spA{\tP_{1,m}}{m+1} \prod_{i=m+1}^{n-1} \spA{i}{i+1}} +
  \mathcal{O}(\smallpar^{3-m})
  \nonumber\\&
  =
  \Sp^{(0)}(-\tP_{1,m}^+;1^-,2^+,\ldots,m^+) \,
  A^{(0)}_{n-m+1}(\tP_{1,m}^-,(m+1)^+,\ldots,r^-,\ldots,n^+)
  + \mathcal{O}(\smallpar^{3-m})
\end{align}
where we have used eq. \eqref{eq:kTdef} to perform the power counting. For $i,j \in [1,m]$
this can be seen explicitly,
\begin{align}
  \wspA{i} &= -\smallpar\frac{\spAB{\tP_{1,m}}{k_T}{\eta}}{2(P_{1,m}\cdot\eta)\zspB{j}} = \mathcal{O}(\smallpar)
  \nonumber\\
  \Rightarrow
  \spA{i}{j} &= \left( \zspA{i}\wspA{j}-\zspA{j}\wspA{i} \right) \spA{\tP_{1,m}}{\eta} =
  \mathcal{O}(\smallpar).
\end{align}
One can clearly arrive at this final result without being so explicit about the parametrisation,
yet it is convenient to have one in a generic implementation.

\subsection{One-loop basis functions for $pp\to H+2j$ in the triple collinear limit}

The analytic $H+4$ parton amplitudes have been computed using unitarity cuts and expressed
in terms of the universal infrared poles plus finite logarithmic and di-logarithmic
functions as well as rational terms. Taking the triple collinear limit of the infrared poles, rational terms and logarithms
as above presents no difficulties. Dealing with the di-logarithmic parts requires some minor effort
to ensure the arguments are in the appropriate region so the limit will converge. Polylogarithmic
identities are well known and understood in huge detail (see Ref. \cite{Duhr:2014woa} for a recent review) - way beyond the simple structures appearing
here. Nevertheless we collect some potentially useful identities here to aid the reader,
\begin{align}
  &\text{Li}_2\left( 1-x \right) + \text{Li}_2\left( x \right) + \log(x)\log(1-x) - \frac{\pi^2}{6} =
  0  & x\in[0,1] \\
  &\text{Li}_2\left( x \right) + \text{Li}_2\left( \frac{1}{x} \right) + \frac{1}{2}\log(-x)^2 + \frac{\pi^2}{6} =
  0   & x<0 \\
  &\text{Li}_{2}\left(\frac{xy}{(1-x)(1-y)}\right)  - \text{Li}_{2}\left(-\frac{x}{1-x}\right) - \text{Li}_{2}\left(-\frac{y}{1-y}\right)+  \notag \\
  & \qquad  - \text{Li}_{2}\left(\frac{x}{1-y}\right)  - \text{Li}_{2}\left(\frac{y}{1-x}\right) -\text{log}^{2}\left(\frac{1-x}{1-y}\right) =
  0   & x,y\in[0,1]
  \label{eq:dilogids}
\end{align}
One function requiring a bit more thought is the three mass triangle which has square roots
appearing in the arguments of the di-logarithms~\cite{Lu:1992ny,Bern:1993kr,Binoth:2001vma,vanHameren:2005ed,Ellis:2007qk}:
\begin{align}
  \text{I}_3^{3m}\left(s_{ij},s_{kl},m_H^2\right) 
  \xrightarrow[]{i||j||k} \ &
  \frac{1}{\left(1-z_k\right) m_H^2}\Biggl( \text{Li}_2\left(1-z_k\right)-\text{Li}_2\left(1-\frac{1}{z_k}\right)\notag \\ & 
  -\frac{1}{2}\log ^2\left(z_k\right)- \log \left(\frac{m_H^2}{s_{ij}}\right) \log\left(z_k\right)\Biggr)
\end{align}
One other minor issue with the results available in the literature is that the
NMHV expressions have been presented using Forde's method for triple cuts \cite{Forde:2007mi}. This method
gives the coefficients as the sum over solutions to the on-shell equations. To aid our computation
we performed this sum explicitly to write the coefficients in terms of the usual spinor products of
the external momenta.

\section{$g\to ggg$ splitting amplitudes}
\subsection{Colour structure and primitive amplitude decomposition}

In the section we will suppress all helicity superscripts and the function arguments
are taken to represent both momenta and helicity. The tree-level colour decomposition can be written as,
\begin{align}
  \cSp^{(0)}&(\{a_P,a_1,a_2,a_3\},-P; 1, 2, 3)
  \nonumber\\
  &= \sum_{\sigma \in S_3} \trF(a_P, a_{\sigma(1)}, a_{\sigma(2)}, a_{\sigma(3)})
    \Sp^{(0)}(-P; \sigma(1), \sigma(2), \sigma(3)) \\
  &=
  \sum_{\sigma \in S_2} \tf^{a_1 a_{\sigma(2)} b} \tf^{b a_{\sigma(3)} a_P}
    \Sp^{(0)}(-P; 1, \sigma(2), \sigma(3))
\end{align}
where $\trF(a_1,\dots,a_n) = T^{a_1}_{j i_1}T^{a_2}_{i_1 i_2}\dots T^{a_n}_{i_{n-1} j}$ in terms of the fundamental
generators of $SU(N_c)$ and $\tf^{abc} = i\sqrt{2}f^{abc}$ in terms of the adjoint structure
constants. The relation between the two representations can be shown to hold using the Kleiss-Kuijf
relations \cite{Kleiss:1988ne} for the splitting amplitudes,
\begin{align}
  \Sp^{(0)}(-P; 3, 2, 1) &=  \Sp^{(0)}(-P; 1, 2, 3) \\
  \Sp^{(0)}(-P; 1, 3, 2) &= -\Sp^{(0)}(-P; 1, 2, 3)-\Sp^{(0)}(-P; 1, 3, 2)
 \label{eq:KKrelation}
\end{align}
The one-loop colour decomposition is\footnote{We write the one-loop decomposition in the standard trace basis rather than the
slightly more compact `F-basis' representation of Del Duca-Maltoni-Dixon \cite{DelDuca:1999rs}.
Since we express the colour summed squared matrix element in terms of the minimal basis of primitive amplitudes the
final expressions are equivalent to the DDM forms.},
\begin{align}
  \cSp^{(1)}&(\{a_P,a_1,a_2,a_3\},-P; 1, 2, 3)
  \nonumber\\
  &= \sum_{\sigma \in S_3} \trF(a_P, a_{\sigma(1)}, a_{\sigma(2)},a_{\sigma(3)})
    \Sp^{(1)}_{1}(-P; \sigma(1), \sigma(2), \sigma(3))
  \nonumber\\&
  + \sum_{\sigma \in S_3/Z_2} \trF(a_P, a_{\sigma(1)} \trF(a_{\sigma(2)},a_{\sigma(3)})
    \Sp^{(1)}_{3}(-P; \sigma(1), \sigma(2), \sigma(3))
\end{align}
where the partial amplitudes are composed of primitive amplitudes as follows:
\begin{align}
  &\Sp^{(1)}_{1}(-P; 1, 2, 3) \nonumber\\
  &= N_c \Sp^{[g]}(-P; 1, 2, 3) - N_f \Sp^{[f]}(-P;
  1, 2, 3), \\
  &\Sp^{(1)}_{3}(-P; 1, 2, 3) \nonumber\\
  &= 2 \left( \Sp^{[g]}(-P; 1, 2, 3)
         +\Sp^{[g]}(-P; 1, 3, 2)
         +\Sp^{[g]}(-P; 3, 1, 2)\right).
\end{align}
The primitive amplitudes for the gluon and fermion loops obey line-reversal
symmetry,
\begin{equation}
  \Sp^{[X]}(-P; 1, 2, 3)
  = \Sp^{[X]}(-P; 3, 2, 1)
  \label{eq:lrev}
\end{equation}
and so in all we have three independent gluon loop primitive amplitudes, three fermion loop
primitive amplitudes and two tree-level primitive amplitudes. The colour summed Born and virtual
corrections can then be written according to \eqref{eq:csumfactprimitive} using:
\begin{align}
  \vec{\Sp}^{[0]} &= \begin{pmatrix} \Sp^{[0]}(-P; 1, 2, 3) \\ \Sp^{[0]}(-P; 1, 3, 2)\end{pmatrix} \\
  \mathcal{C}_{\Sp}^{[0,0]} &= N_c^2 \begin{pmatrix} 4 & 2 \\ 2 & 4 \end{pmatrix} \\
    \vec{\Sp}^{[1]} &=  \begin{pmatrix}
    N_c\, \Sp^{[g]}(-P; 1, 2, 3) \\
    N_c\, \Sp^{[g]}(-P; 2, 1, 3) \\
    N_c\, \Sp^{[g]}(-P; 2, 3, 1) \\
    N_f\, \Sp^{[f]}(-P; 1, 2, 3) \\
    N_f\, \Sp^{[f]}(-P; 2, 1, 3) \\
    N_f\, \Sp^{[f]}(-P; 2, 3, 1)
  \end{pmatrix} \\
    \mathcal{C}_{\Sp}^{[0,1]} &= 2 N_c^2 \begin{pmatrix}
      2 & -2 & 0 & -2 & 2 & 0 \\
      0 & -2 & 2 & 0 & 2 & -2 \\
    \end{pmatrix}
\end{align}

We also choose to present the results using the super-symmetric decomposition:
\begin{align}
  &\Sp^{[g]}(-P; 1, 2, 3) \nonumber\\ &=
    \Sp^{[\cN=4]}(-P; 1, 2, 3)
  + 4 \Sp^{[\cN=1]}(-P; 1, 2, 3)
  + (1 - \eps\delta_R)\Sp^{[\cN=0]}(-P; 1, 2, 3) \\
  &\Sp^{[f]}(-P; 1, 2, 3) \nonumber\\ &=
    \Sp^{[\cN=1]}(-P; 1, 2, 3)
  + \Sp^{[\cN=0]}(-P; 1, 2, 3)
  \label{eq:SUSYdecomp}
\end{align}
since this yields particularly compact expressions. We also include the scheme dependence for both
the FDH ($\delta_R=0$) and CDR ($\delta_R=1$) schemes.

\subsection{Results}

We define the following phase-free quantities,
\begin{align}
  \spr_{ij} \equiv \spr_{ijk} &= \frac{\spA{i}{j}\zspA{k}}{\spA{j}{k}\zspA{i}}, &
  \sptr_{ij} \equiv \sptr_{ijk} &= \frac{\spB{i}{j}\wspB{k}}{\spB{j}{k}\wspB{i}}, &
  \spX_{ij} &= \frac{\zspA{i}\spB{ij}}{\spB{j}{\tilde{P}}}.
  \label{eq:ratiodefs}
\end{align}
Since there can be no repeated index in either $\spr_{ijk}$ and $\sptr_{ijk}$ each can be uniquely
specified by the two first labels.

The integral functions are defined using the following basis,
\begin{align}
F^{\text{MHV}} &=
 \frac{1}{2} \left(
  \log ^2\left(z_1\right) 
+ \log ^2\left(z_3\right)
+ \frac{\pi ^2}{3}\right)
-\log \left( \frac{s_{12}}{s_{123}} \right)\log \left( \frac{s_{23}}{s_{123}} \right) \nonumber\\&
+\log \left(\frac{1-z_3}{z_1}\right) \log \left(\frac{s_{12}}{s_{123}}\right)
+\log \left(\frac{1-z_1}{z_3}\right) \log \left(\frac{s_{23}}{s_{123}}\right)
\nonumber\\&
+\text{Li}_2\left(-\frac{z_2}{z_1}\right)
+\text{Li}_2\left(-\frac{z_2}{z_3}\right)
+\text{Li}_2\left(-\frac{z_3}{1-z_3}\right)
+\text{Li}_2\left(-\frac{z_1}{1-z_1}\right)
\nonumber\\&
-\text{Li}_2\left(1-\frac{s_{12}}{\left(1-z_3\right) s_{123}}\right)
-\text{Li}_2\left(1-\frac{s_{23}}{\left(1-z_1\right) s_{123}}\right) \\
F_1^{\text{NMHV}} &=
-\log \left(1-z_3\right) \left( \log \left(\frac{z_1 z_3}{1-z_3}\right) + \log \left(\frac{s_{12}}{s_{23}}\right) \right)
+\log \left(z_1 z_3\right) \log \left(\frac{s_{12}}{s_{123}}\right)
\nonumber\\&
-\frac{1}{2} \left(\log \left(z_3\right) \log \left(\frac{s_{12}}{s_{123}}\right)+\log
\left(z_1\right) \log \left(\frac{s_{23}}{s_{123}}\right)-\frac{\pi ^2}{3}\right)\\
F_2^{\text{NMHV}} &= F_1^{\text{NMHV}} \big|_{1\leftrightarrow 3}\\
F_3^{\text{NMHV}} &=
\frac{1}{2} \left(\log \left(z_3\right) \log \left(\frac{s_{12}}{s_{123}}\right)+\log \left(z_1\right) \log \left(\frac{s_{23}}{s_{123}}\right)-\frac{\pi ^2}{3}\right) \nonumber\\&
-\log \left(\frac{s_{12}}{s_{123}}\right) \log \left(\frac{s_{23}}{s_{123}}\right)\\
F_{\text{box}}^{\text{1m}} &= - \frac{\pi^2}{3} - \log^2\left(\frac{s_{12}}{s_{23}}\right) - 2 \left( \text{Li}_2\left(1-\frac{s_{123}}{s_{12}}\right)
+\text{Li}_2\left(1-\frac{s_{123}}{s_{23}}\right) \right) \\
\hat{L}_0\left(s_{1},s_{2}\right) & = \log\left(\frac{s_1}{s_2}\right) \\
\hat{L}_1\left(s_{1},s_{2}\right) & = \frac{1}{s_1-s_2}\, \log\left(\frac{s_1}{s_2}\right) \\
\hat{L}_2\left(s_{1},s_{2}\right) & = \frac{1}{(s_1-s_2)^2}\, \log\left(\frac{s_1}{s_2}\right) - \frac{1}{2} \frac{1}{s_1-s_2}\left(\frac{1}{s_1}+\frac{1}{s_2}\right) \\
\hat{L}_3\left(s_{1},s_{2}\right) & = \frac{1}{(s_1-s_2)^3}\, \log\left(\frac{s_1}{s_2}\right) - \frac{1}{2} \frac{1}{(s_1-s_2)^2}\left(\frac{1}{s_1}+\frac{1}{s_2}\right)
  \label{eq:SpLogs}
\end{align}
We express the infrared poles and associated logarithms as described by Catani's formula~\cite{Catani:2003vu},
\begin{align}
  V_g = -\frac{1}{\epsilon ^2}\left(
    \left(\frac{\mu _R}{-s_{12}}\right)^{\epsilon}
  + \left(\frac{\mu _R}{-s_{23}}\right)^{\epsilon}
  + \left(\frac{\mu _R}{-s_{123}}\right)^{\epsilon } \left(z_1{}^{-\epsilon }+z_3{}^{-\epsilon }-2\right)
  \right)
\end{align}
All results in this section are presented unrenormalized.

The tree-level splitting amplitudes are,
\begin{align}
\Sp^{(0)}\left(-P^+;1^+,2^+,3^+\right) &=
0
\\
\Sp^{(0)}\left(-P^+;1^-,2^-,3^-\right) &=
\frac{1}{\left[z_1\right] \left[z_3\right] [12] [23]}
\\
\Sp^{(0)}\left(-P^+;1^+,2^+,3^-\right) &=
\frac{\la z_3\ra {}^3}{\la z_1\ra  \la 12\ra  \la 23\ra }
\\
\Sp^{(0)}\left(-P^+;1^+,2^-,3^+\right) &=
\frac{\la z_2\ra {}^4}{\la z_1\ra  \la z_3\ra  \la 12\ra  \la 23\ra }
\\
\Sp^{(0)}\left(-P^+;1^+,2^-,3^-\right) &=
-\frac{[1P]^2}{[23]^2}
 \left(
\frac{\sptr_{32}}{s_{123}}+\frac{\spX_{23}^3 \spr_{12} \sptr_{12}^2 s_{23}}{\left(1-z_3\right) z_3 s_{12}^2}
\right)
\\
\Sp^{(0)}\left(-P^+;1^-,2^+,3^-\right) &=
-\frac{[2P]^2}{[13]^2 s_{1P} s_{3P}} \Bigg(\nonumber\\
\frac{\sptr_{21} \sptr_{23} s_{13}^4}{s_{12} s_{123} s_{23}}&
+\frac{\spX_{32}^2 z_2 z_3 s_{1P} \left(\spr_{13} \spr_{31}\right){}^{\dagger }}{\spX_{23} \left(1-z_3\right)}
+\frac{\spX_{12}^2 z_1 z_2 s_{3P} \left(\spr_{13} \spr_{31}\right){}^{\dagger }}{\spX_{21} \left(1-z_1\right)}
\Bigg)
  \label{eq:g2gggTree}
\end{align}
All other helicity configurations are given via parity or the line-reversal symmetry of eq. \eqref{eq:lrev}.
The one-loop splitting primitive amplitudes are,
\begin{align}
\Sp&^{[\cN=4]}\left(-P^+; 1^+, 2^+, 3^+\right) = 0 \\
\Sp&^{[\cN=4]}\left(-P^+; 1^-, 2^-, 3^-\right) = \Sp^{(0)}\left(-P^+; 1^-, 2^-, 3^-\right)
\left( V_g + F^{\text{MHV}} \right) \\
\Sp&^{[\cN=4]}\left(-P^+; 1^+, 2^+, 3^-\right) = \Sp^{(0)}\left(-P^+; 1^+, 2^+, 3^-\right)
\left( V_g + F^{\text{MHV}} \right) \\
\Sp&^{[\cN=4]}\left(-P^+; 1^+, 2^-, 3^+\right) = \Sp^{(0)}\left(-P^+; 1^+, 2^-, 3^+\right)
\left( V_g + F^{\text{MHV}} \right) \\
\Sp&^{[\cN=4]}\left(-P^+; 1^+, 2^-, 3^-\right) = \Sp^{(0)}\left(-P^+; 1^+, 2^-, 3^-\right)
\left(V_g\right)
-\frac{[1P]^2}{[23]^2}
\Bigg(
\nonumber\\&
+ \left(
\frac{\spX_{23} \spr_{12} \sptr_{12} s_{23}}{z_3 s_{12}^2} \left(\frac{s_{12} \left(z_1-1\right){}^3}{z_2 s_{1P}}+\frac{2 \spX_{23}^2 \sptr_{12}}{z_3-1}\right)
+\frac{1}{s_{123}}\left( \frac{\sptr_{12} s_{23}^3}{s_{12} s_{1P} s_{3P}}-\sptr_{32}\right) \right) F_1^{\text{NMHV}}
\nonumber\\&
-\frac{\spX_{23} \spr_{12} \sptr_{12} s_{23} \left(z_1-1\right){}^3}{z_2 z_3 s_{12} s_{1P}} F_2^{\text{NMHV}}
\nonumber\\&
+\frac{1}{s_{123}}\left(\frac{\sptr_{12} s_{23}^3}{s_{12} s_{1P} s_{3P}}+\sptr_{32}\right) F_3^{\text{NMHV}}
\Bigg)
\\
\Sp&^{[\cN=4]}\left(-P^+; 1^-, 2^+, 3^-\right) = \Sp^{(0)}\left(-P^+; 1^-, 2^+, 3^-\right)
\left(V_g\right)
-\frac{[2P]^2}{[13]^2}
\Bigg(
\nonumber\\&
+ \left(\frac{\spX_{32} \sptr_{31} s_{13} \spX_{12}^2}{\spX_{23} z_3 s_{12} s_{3P}}
+\frac{z_2 z_3 \left(\spr_{13} \spr_{31}\right){}^{\dagger } \spX_{32}^2}{\spX_{23} s_{3P}(z_3-1)}
+\frac{\spX_{12}^2 z_2 \sptr_{31}^2 \spX_{32}^2}{\spX_{23}^2 z_3 s_{3P}(z_3-1)}
\right) F_1^{\text{NMHV}}
\nonumber\\&
+ \left(\frac{z_1 \left(\spr_{13} \spr_{31}\right){}^{\dagger } \spX_{12}^2}{s_{1P}}
+ \frac{\spX_{12} \sptr_{13} \sptr_{31} \spX_{32}^3}{\spX_{23} z_1 s_{23} (z_1-1)}
+ \frac{\spX_{12} z_1{}^2 \left(\spr_{13} \spr_{31}\right){}^{\dagger } \spX_{32}}{s_{1P} (z_1-1)}
\right) F_2^{\text{NMHV}}
\nonumber\\&
+ \frac{1}{s_{123}} \left(\frac{\sptr_{21} \sptr_{23} s_{13}^4}{s_{12} s_{23} s_{1P} s_{3P}}+\sptr_{13} \sptr_{31}\right) F_3^{\text{NMHV}}
\Bigg)
  \label{g2gggN4}
\end{align}

\begin{align}
\Sp&^{[\cN=1]}\left(-P^+; 1^+, 2^+, 3^+\right) =
0\\
\Sp&^{[\cN=1]}\left(-P^+; 1^-, 2^-, 3^-\right) =
0\\
\Sp&^{[\cN=1]}\left(-P^+; 1^+, 2^+, 3^-\right) = \Sp^{(0)}\left(-P^+; 1^+, 2^+, 3^-\right)
\spr_{32} s_{12} \hat{L}_1\left(s_{23},s_{123}\right)  \\
\Sp&^{[\cN=1]}\left(-P^+; 1^+, 2^-, 3^+\right) = \Sp^{(0)}\left(-P^+; 1^+, 2^-, 3^+\right)
\frac{\spr_{23}}{\spr_{31}}
\Bigg(\nonumber\\&\frac{F_{\text{box}}^{\text{1m}}}{2}-\left(\hat{L}_1\left(s_{12},s_{123}\right)+\hat{L}_1\left(s_{23},s_{123}\right)\Bigg) s_{13}
\right)\\
\Sp&^{[\cN=1]}\left(-P^+; 1^+, 2^-, 3^-\right) =
-\frac{[1P]^2 s_{23}}{[23]^2 s_{123}} \hat{L}_1\left(s_{12},s_{123}\right) \\
\Sp&^{[\cN=1]}\left(-P^+; 1^-, 2^+, 3^-\right) =
-\frac{[2P]^2}{[13]^2 s_{123}} \Bigg(\nonumber\\&
\frac{F_{\text{box}}^{\text{1m}}}{2}-\left(\hat{L}_1\left(s_{12},s_{123}\right)+\hat{L}_1\left(s_{23},s_{123}\right)\right)
s_{13}\Bigg)
  \label{eq:g2gggN1}
\end{align}

\begin{align}
\Sp&^{[\cN=0]}\left(-P^+; 1^+, 2^+, 3^+\right) =
-\frac{[1P] [3P]}{3 \la 12\ra  \la 23\ra }
\left(\frac{1}{s_{123}}-\frac{\spX_{23}^2 \sptr_{13} \left(\spr_{32} s_{12}+\spr_{12} s_{23}\right)}{\sptr_{31} s_{12} s_{23}}\right)
\\
\Sp&^{[\cN=0]}\left(-P^+; 1^-, 2^-, 3^-\right) =
\frac{1}{3} \Sp^{(0)}\left(-P^+; 1^-, 2^-, 3^-\right) \Bigg(\nonumber\\&
z_1 z_2+\frac{z_1 \left(1-z_2{}^2\right) z_3 z_2}{\left(1-z_1\right) \left(1-z_2\right) \left(1-z_3\right)}+z_3 z_2+z_1 z_3
-\frac{z_1 z_3}{\spX_{12} \spX_{32}} \left(\frac{\spX_{32} z_1}{1-z_3}+\frac{\spX_{12} z_3}{1-z_1}+\frac{s_{13}}{s_{123}}\right)
\Bigg)
\\
\Sp&^{[\cN=0]}\left(-P^+; 1^+, 2^+, 3^-\right) =
\Sp^{(0)}\left(-P^+; 1^+, 2^+, 3^-\right) \frac{\spX_{12}  s_{23}} {3 \spX_{32}^3}
\Bigg(\nonumber\\&
\frac{\spX_{32} z_2}{s_{12}} \left(\frac{\spX_{12}}{z_3}+\frac{\spX_{32}}{z_3-1}\right)
-\spX_{12} \spX_{32} \hat{L}_2\left(s_{23},s_{123}\right) s_{23}
+2 \spX_{12} \sptr_{23} \hat{L}_3\left(s_{23},s_{123}\right) s_{13} s_{23} \nonumber\\&
+\frac{1}{s_{123}}\left(\frac{s_{23}}{s_{12}}-\frac{1}{2} \left(\spX_{12}+2\right) \spX_{32}\right)
\Bigg)
\\
\Sp&^{[\cN=0]}\left(-P^+; 1^+, 2^-, 3^+\right) = \Sp^{(0)}\left(-P^+; 1^+, 2^-, 3^+\right)
\frac{\spr_{23}^2 s_{13}}{3 \spr_{31}^2}
\Bigg(\nonumber\\&
-\frac{2 \hat{L}_3\left(s_{12},s_{123}\right) s_{13}^2}{\spr_{32}}
-2 \spr_{32} \hat{L}_3\left(s_{23},s_{123}\right) s_{13}^2
-3 \hat{L}_2\left(s_{23},s_{123}\right) \left(2 s_{12}+3 s_{13}\right)\nonumber\\&
+\frac{5}{2} \left(\frac{2}{s_{123}}+\frac{1}{s_{23}}+\frac{1}{s_{12}}\right)
-3 \hat{L}_2\left(s_{12},s_{123}\right) \left(3 s_{13}+2 s_{23}\right)
-\frac{1}{2 s_{123}}\left(\frac{s_{12}}{s_{23}}+\frac{s_{23}}{s_{12}}\right) \nonumber\\&
+\frac{s_{123}}{s_{12} s_{23}}
-\frac{3 F_{\text{box}}^{\text{1m}}}{s_{13}}
\Bigg)
  \label{eq:g2gggN0mhv}
\end{align}
\begin{align}
\Sp&^{[\cN=0]}\left(-P^+; 1^+, 2^-, 3^-\right) =
\frac{[1P]^2}{3 [23]^2}
\Bigg(\nonumber\\&
\frac{2 \spX_{23}^2 \spr_{21}^2 s_{13}^2 s_{23} \sptr_{12}^2}{s_{12}} \hat{L}_3\left(s_{12},s_{123}\right)
+\frac{\spX_{23}^2 \spr_{21} s_{13} \left(-s_{12}-\spr_{21} s_{13}\right) s_{23} \sptr_{12}^2}{s_{12} s_{123}} \hat{L}_2\left(s_{12},s_{123}\right)
\nonumber\\&
+\frac{\spX_{23}^2 \left(3 \spr_{21} s_{13}-s_{12}\right) s_{23} \sptr_{12}^2}{s_{12} s_{123}} \hat{L}_1\left(s_{12},s_{123}\right)
-\frac{\spX_{23}^2 \spr_{12} s_{23} \sptr_{12}^2}{s_{12}^2}
-\frac{\spX_{23}^2 \spr_{21} s_{13} s_{23} \sptr_{12}^2}{s_{12}^2 s_{123}}
\nonumber\\&
+\frac{1}{s_{123}^2}\frac{\spX_{23} \spr_{21} \sptr_{12}^2 \sptr_{32} s_{13} s_{23}}{2 s_{12}}-\frac{\spX_{23}^2 \spr_{21} \sptr_{12}^2 s_{13} s_{23}}{s_{12}}
\Bigg)
\\
\Sp&^{[\cN=0]}\left(-P^+; 1^-, 2^+, 3^-\right) =
\frac{[2P]^2}{3 [13]^2}
\Bigg(\nonumber\\&
-2 \hat{L}_3\left(s_{23},s_{123}\right) s_{123} s_{13} \spX_{12}^2
+\hat{L}_2\left(s_{23},s_{123}\right) \left(3 \spX_{12} s_{123}+s_{13}\right) \spX_{12}
\nonumber\\&
-2 \spX_{32}^2 \hat{L}_3\left(s_{12},s_{123}\right) s_{123} s_{13}
+\frac{3 \left(2 s_{12}+4 \spX_{12} s_{123}+s_{13}\right)}{s_{123} s_{13}} \hat{L}_0\left(s_{23},s_{123}\right)
\nonumber\\&
+\spX_{32} \hat{L}_2\left(s_{12},s_{123}\right) \left(3 \spX_{32} s_{123}+s_{13}\right)
+\hat{L}_1\left(s_{23},s_{123}\right) \left(\frac{s_{13}}{s_{123}}-\frac{6 \spX_{12}^2 s_{123}}{s_{13}}\right)
\nonumber\\&
+\hat{L}_1\left(s_{12},s_{123}\right) \left(\frac{s_{13}}{s_{123}}-\frac{6 \spX_{32}^2 s_{123}}{s_{13}}\right)
+\frac{3 \left(4 \spX_{32} s_{123}+s_{13}+2 s_{23}\right)}{s_{123} s_{13}} \hat{L}_0\left(s_{12},s_{123}\right)
\nonumber\\&
+\frac{1}{2 s_{12} s_{123} s_{23}}\Big(
-4 s_{123} s_{23} \spX_{32}^2
+\left(\spX_{12} s_{123} \left(-3 s_{12}+2 s_{123}-3 s_{23}\right)+\left(5 s_{12}-s_{23}\right) s_{23}\right) \spX_{32}
\nonumber\\&
-\spX_{12} s_{12} \left(s_{12}+4 \spX_{12} s_{123}-5 s_{23}\right)
\Big)
-\frac{3 \sptr_{21} \sptr_{23}}{s_{123}} F_{\text{box}}^{\text{1m}}
\Bigg)
  \label{eq:g2gggN0nmhv}
\end{align}

\section{$g\to \qb qg$ splitting amplitudes}

\subsection{Colour structure and primitive amplitude decomposition}

\begin{figure}
{\scriptsize
\begin{align*}
  \Sp^{[L]}(-P;1_{\bar{q}},2_q,3) &= \,\,
  \parbox{3cm}{\includegraphics[width=3cm]{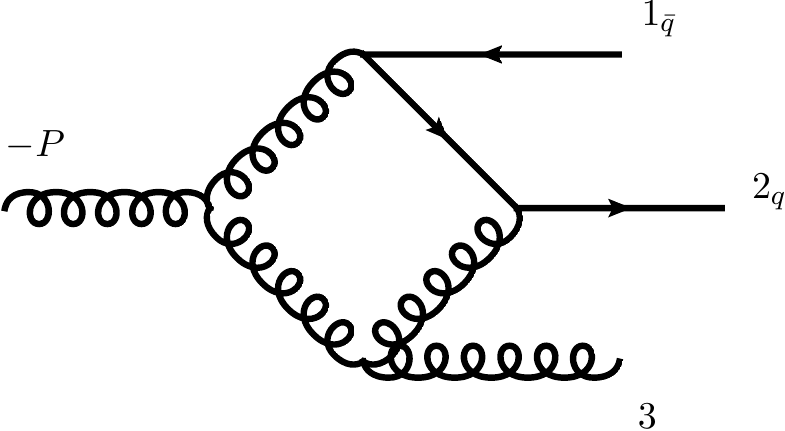}} \,\,
  + \dots \\
  \Sp^{[R]}(-P;1_{\bar{q}},2_q,3) &= \,\,
  \parbox{3cm}{\includegraphics[width=3cm]{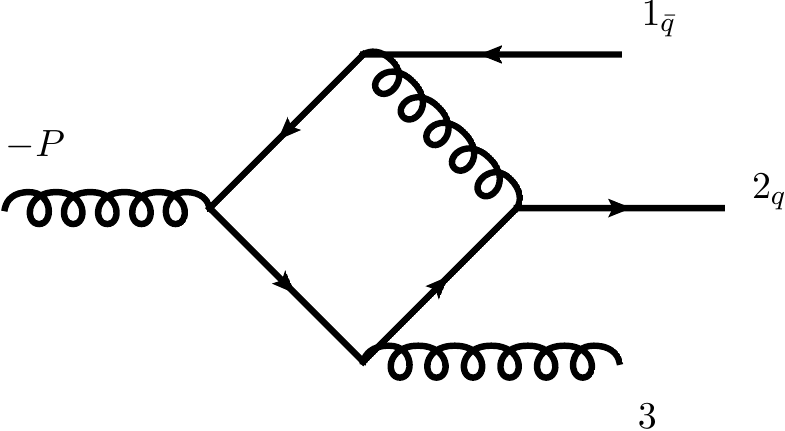}} \,\,
  + \dots \\
  \Sp^{[f]}(-P;1_{\bar{q}},2_q,3) &= \,\,
  \parbox{3cm}{\includegraphics[width=3cm]{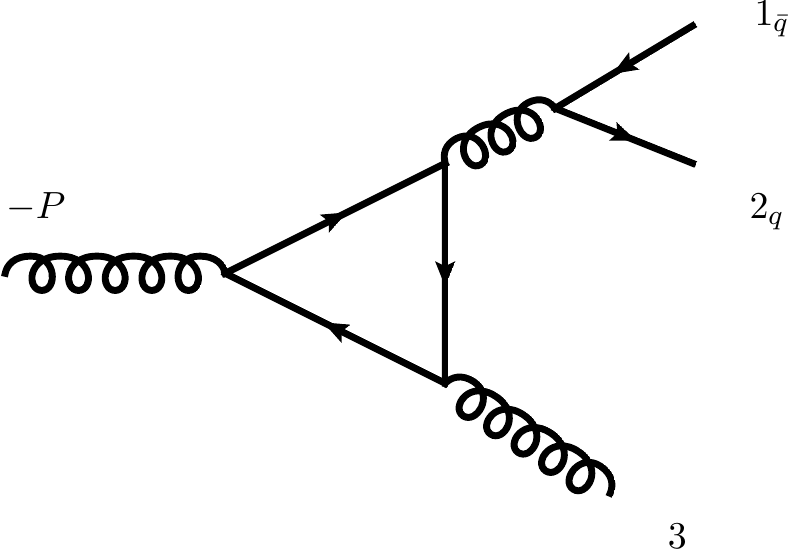}} \,\,
  + \dots \\
  \Sp^{[L+R]}(-P;1_{\bar{q}},2,3_q) &= \,\,
    \parbox{3cm}{\includegraphics[width=3cm]{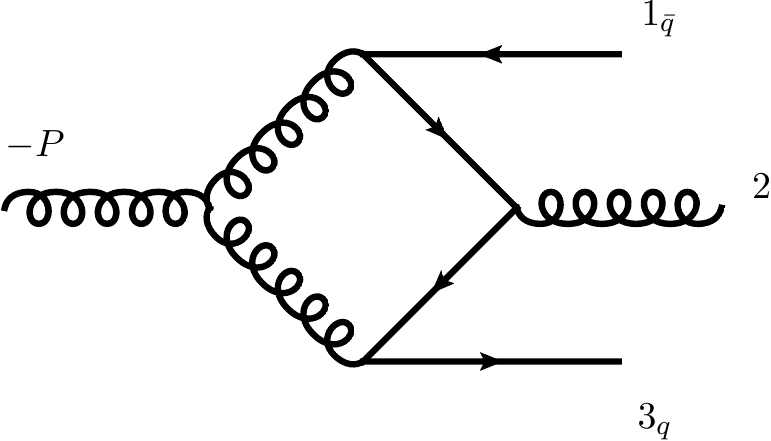}} \,\,
  + \parbox{3cm}{\includegraphics[width=3cm]{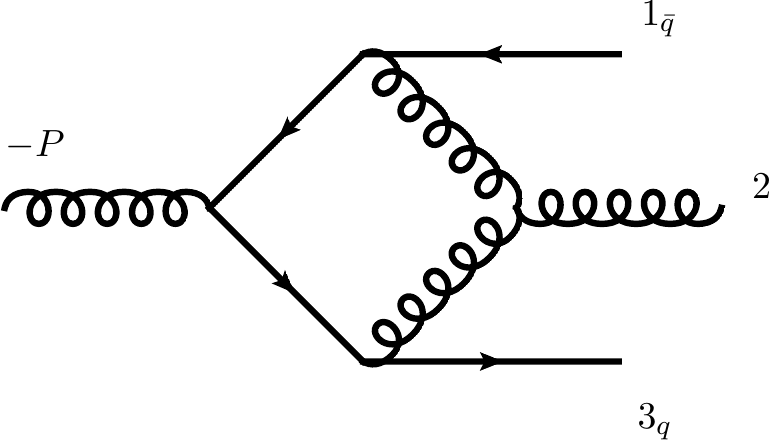}} \,\,
  + \dots 
\end{align*}
}
\caption{Sample diagrams corresponding to primitive amplitudes for $\Sp^{(1)}(-P;1_{\bar{q}}, 2_q, 3_g)$}
\label{fig:exampleprim}
\end{figure}

The colour structure of the tree-level splitting amplitudes is
\begin{align}
  \cSp^{(0)}&(\{a_P,\bar{\imath}_1,i_2,a_3\},-P, 1_{\bar{q}}, 2_q, 3) = \nonumber\\&
   T\left(a_P, a_3\right)^{\bar{\imath}_1}_{i_2} \Sp^{(0)}(-P;1_{\bar{q}},2_q,3)\;+\;
    T\left(a_3, a_P\right)^{\bar{\imath}_1}_{i_2} \Sp^{(0)}(-P;2_q,1_{\bar{q}},3)
\end{align}
where $T(a_1,\dots,a_n)_{i}^{\bar{\jmath}} = T^{a_1}_{\bar{\jmath} k_1}T^{a_1}_{k_1 k_2}\dots
T^{a_n}_{k_{n-1} i}$. Note that charge conjugation symmetry allows us to write
$\Sp^{(0)}(-P;2_q,1_{\bar{q}},3) = \Sp^{(0)}(-P;2_{\bar{q}},1_q,3)$.
At one-loop we have three colour structures,
\begin{align}
  \cSp^{(1)}(\{a_P,\bar{\imath}_1,i_2,a_3\},-P; 1_{\bar{q}}, 2_q, 3) ={} &
      N_c\ \Big[T\left(a_P, a_3\right)^{\bar{\imath}_1}_{i_2} \ \Sp^{(1)}_{4;1}(-P; 1_{\bar{q}}, 2_q, 3)  \notag \\
	 & \qquad  +T\left(a_3, a_P\right)^{\bar{\imath}_1}_{i_2} \ \Sp^{(1)}_{4;1}(-P; 2_q, 1_{\bar{q}}, 3) \Big]  \notag \\
      & + \delta^{a_Pa_3}\delta^{\bar{\imath}_1}_{i_2} \ \Sp^{(1)}_{4;3}(-P; 1_{\bar{q}}, 2_q, 3).
\end{align}
The partial amplitudes $\Sp_{4;1}$ and $\Sp_{4;3}$ are given in terms of the primitive amplitudes
\begin{align}
 \Sp_{4;1}(-P; 1_{\bar{q}}, 2_q, 3) &= \Sp^{[L]}(-P; 1_{\bar{q}}, 2_q, 3) - \frac{1}{N_c^2}\Sp^{[R]}(-P; 1_{\bar{q}}, 2_q, 3) \notag \\
				    &+ \frac{n_f}{N_c}\Sp^{[f]}(-P; 1_{\bar{q}}, 2_q, 3) \\[0.3in]
 \Sp_{4;3}(-P; 1_{\bar{q}}, 2_q, 3) &= \Sp^{[L]}(-P; 1_{\bar{q}}, 2_q, 3) + \Sp^{[L]}(-P; 2_{\bar{q}}, 1_q, 3)
					+ \Sp^{[L+R]}(-P; 1_{\bar{q}}, 3, 2_q) \notag \\
				    & + \Sp^{[R]}(-P; 1_{\bar{q}}, 2_q, 3)
					+ \Sp^{[R]}(-P; 2_{\bar{q}}, 1_q, 3)
\end{align}
where the indices $[L]$ and $[R]$ label the primitive amplitudes corresponding to fermion lines
turning left or right upon entering the loop and $[f]$ denotes the primitive amplitudes with
fermion-loop contribute. The label $[L+R]$ in the sub-leading colour
amplitude corresponds to the sum of the left and right primitive amplitudes for the non-adjacent
fermion configuration.
Some representative diagrams of the primitive amplitudes are depicted in fig.~\ref{fig:exampleprim}.

The colour summed Born and virtual corrections can be written as in Eq.~\eqref{eq:csumfact}, where the
vectors $\vec{\Sp}^{(L)}$ and the colour matrices
$\mathcal{C}_{\Sp}^{(L,L')}$ are now given by
\begin{align}
  \vec{\Sp}^{(0)} &= \begin{pmatrix} \Sp^{(0)}(-P; 1, 2, 3) \\ -\Sp^{(0)}(-P; 2, 1, 3)\end{pmatrix} \\
  \mathcal{C}_{\Sp}^{(0,0)} &= \frac{1}{N_c} \begin{pmatrix} N_c^2-1 & -1 \\ -1 & N_c^2-1 \end{pmatrix} \\
  \vec{\Sp}^{(1)} &= \begin{pmatrix} N_c\, \Sp_{4;1}(-P; 1, 2, 3) \\
    -N_c\, \Sp_{4;1}(-P; 2, 1, 3) \\
    \Sp_{4;3}(-P; 1, 2, 3)
  \end{pmatrix} \\
  \mathcal{C}_{\Sp}^{(0,1)} &= \frac{1}{N_c} \begin{pmatrix} N_c^2-1 & -1 & N_c \\ -1 & N_c^2-1 & N_c \end{pmatrix} \\
\end{align}

\begin{figure}
{\scriptsize
\begin{align*}
  \Sp^{[\text{scalar}]}(-P;1_{\bar{q}},2_q,3) &= \,\,
    \parbox{3cm}{\includegraphics[width=3cm]{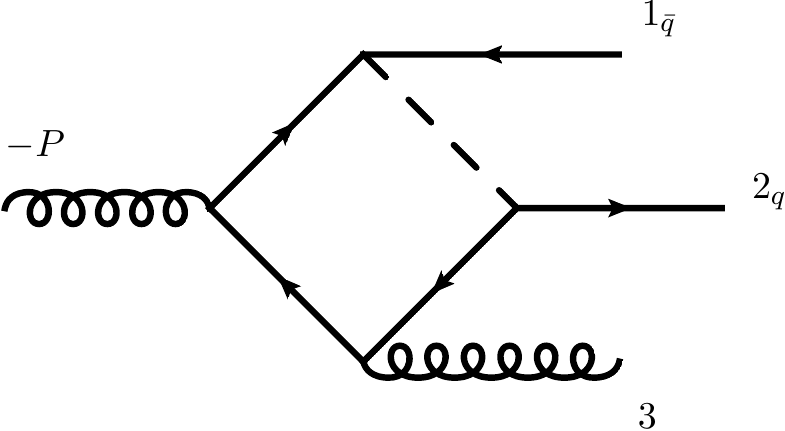}} \,\,
  + \parbox{3cm}{\includegraphics[width=3cm]{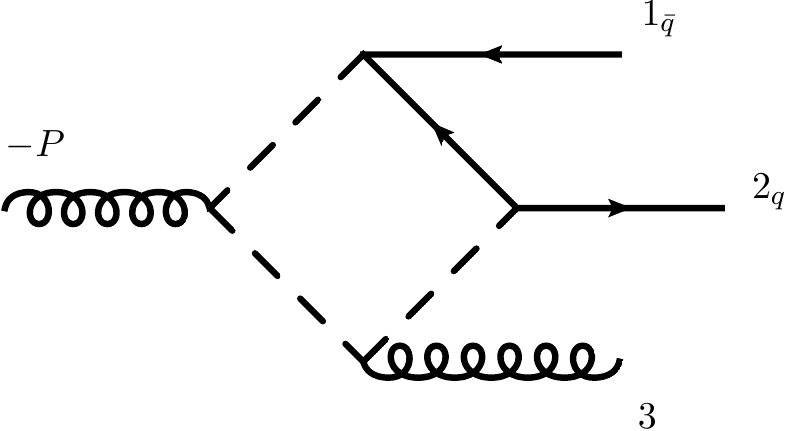}} \,\,
  + \parbox{3cm}{\includegraphics[width=3cm]{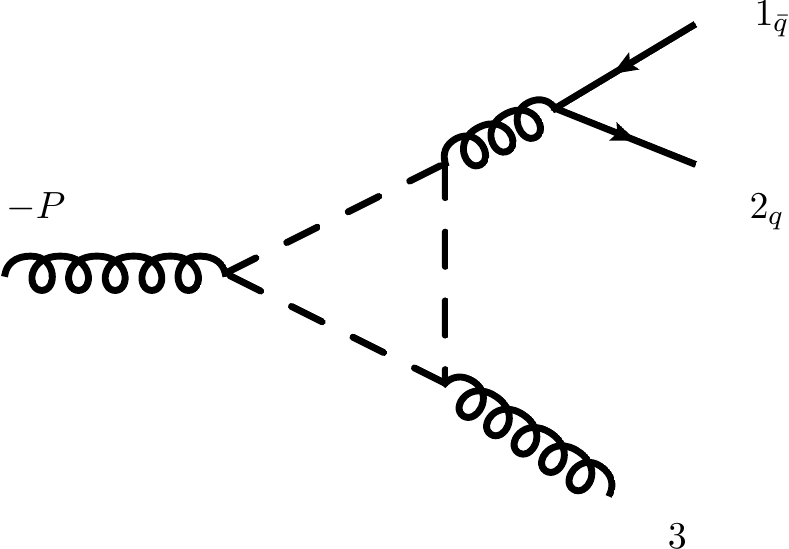}} \,\,
  + \dots \\
  \Sp^{[\text{scalar}]}(-P;1_{\bar{q}},2,3_q) &= \,\,
    \parbox{3cm}{\includegraphics[width=3cm]{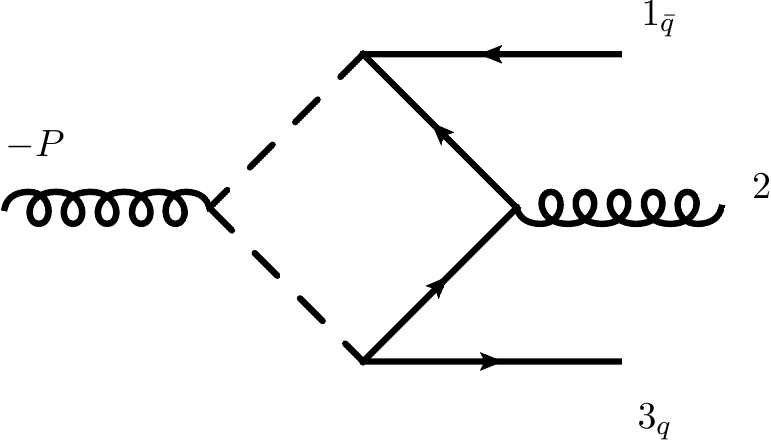}} \,\,
  + \parbox{3cm}{\includegraphics[width=3cm]{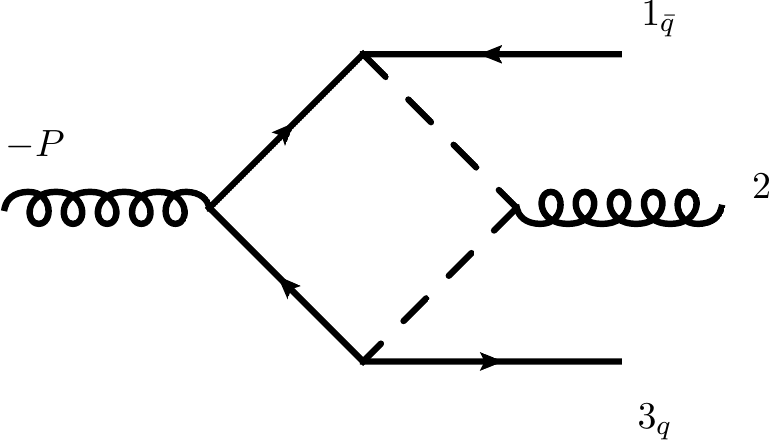}} \,\,
  + \dots 
\end{align*}
}
\caption{Sample diagrams corresponding to scalar contribution for $\Sp^{(1)}(-P;1_{\bar{q}}, 2_q,
3_g)$ in $\cN=4$ super-symmetric Yang-Mills theory.}
\label{fig:examplescalar}
\end{figure}

The quark primitive splitting amplitudes also have a useful super-symmetric decomposition \cite{Bern:1994fz}.
In this case we can write the complicated ``left-moving" amplitudes in terms of simpler ones built using the $\cN=4$ super-multiplet,
\begin{align}
  \Sp^{[L]}(-P; 1_{\qb}, 2_q, 3) ={}&
  \Sp^{[\cN=4]}(-P; 1_{\qb}, 2_q, 3)
  -\Sp^{[R]}(-P; 1_{\qb}, 2_q, 3) \nonumber\\&
  -\Sp^{[f]}(-P; 1_{\qb}, 2_q, 3)
  -\Sp^{[\text{scalar}]}(-P; 1_{\qb}, 2_q, 3) \\
  \Sp^{[L+R]}(-P; 1_{\qb}, 2, 3_q) ={}&
  \Sp^{[\cN=4]}(-P; 1_{\qb}, 2, 3_q)
  -\Sp^{[\text{scalar}]}(-P; 1_{\qb}, 2, 3_q)
\end{align}
where the $\Sp^{[\text{scalar}]}$ function indicates the contribution from the complete scalar
sector of $\cN=4$. This includes a scalar-fermion-fermion vertex as well as the scalar-gluon-gluon vertex
which contributes to the function $\Sp^{[\cN=0]}$ in the pure gluonic case.
Representative diagrams contributing to $\Sp^{[\text{scalar}]}$ are shown in fig.~\ref{fig:examplescalar}. 
Using these
relations we find a compact form for the colour dressed splitting amplitudes in terms of
$\Sp^{[\cN=4]}(-P; 1_{\qb}, 2_q, 3) $,
$\Sp^{[R]}(-P; 1_{\qb}, 2_q, 3)$,
$\Sp^{[f]}(-P; 1_{\qb}, 2_q, 3)$,
$\Sp^{[\text{scalar}]}(-P; 1_{\qb}, 2_q, 3)$,
$\Sp^{[\cN=4]}(-P; 1_{\qb}, 2_q, 3)$ and $\Sp^{[\text{scalar}]}(-P; 1_{\qb}, 2, 3_q)$.

\subsection{Results}

As before all results in this section are presented unrenormalized. The non-vanishing
independent tree-level splitting amplitudes $g \to \bar{q}qg$ are
\begin{align}
\Sp^{(0)}\left(-P^+;1_{\bar{q}}^+,2_{q}^-,3^-\right) &=  -\frac{\la z_2\ra  \la z_3\ra}{\sptr_{12} [23] \la 12\ra }  
							  \left(\frac{\spX_{21}}{\left(1-z_3\right) z_3}+\frac{s_{12}}{\spX_{32} s_{123} \spX_{21}}\right) \\[0.2in]
%
\Sp^{(0)}\left(-P^+;1_{\bar{q}}^-,2_{q}^+,3^-\right) &= -\frac{\la z_3\ra \left( \sptr_{31} \right)^\dagger}{[13] \la 2P\ra } 
							  \Biggl( \nonumber \\& \quad \ 
							  \frac{s_{23} \left(\spr_{13}^2 \sptr_{13}\right)^{\dagger } z_1^2}{ z_3 s_{13} (\sptr_{31})^\dagger}+
							  \frac{s_{13} \left(\spX_{31} \right)^{\dagger }}{z_3 s_{123}}+
							  \frac{z_3^2 s_{12} \left(\spr_{31}^2 \right)^{\dagger }}{\left(1-z_3\right) s_{13}}\Biggr)\\[0.2in]
%
\Sp^{(0)}\left(-P^+;1_{\bar{q}}^+,2_{q}^-,3^+\right) &=  \frac{\la z_2\ra ^3}{\la z_3\ra  \la 12\ra  \la 23\ra }\\[0.2in]
%
\Sp^{(0)}\left(-P^+;1_{\bar{q}}^-,2_{q}^+,3^+\right) & =  -\frac{\la z_1\ra ^2 \la z_2\ra }{\la z_3\ra  \la 12\ra  \la 23\ra }
\label{eqn:treesplitQq}
\end{align}
and the others are obtained by conjugation using the relation
\begin{equation}
 \Sp \left(-P^+;1_{\bar{q}}^{h_1},2_{q}^{h_2},3^{h_3}\right) = 
  \Sp \left(-P^-;1_{\bar{q}}^{-h_1},2_{q}^{-h_2},3^{-h_3} \right) \Big|_{\la ij \ra \leftrightarrow [ij] }  .
\end{equation}
The sub-leading colour tree-level splitting amplitudes $g \to \bar{q}gq$ are not independent because they can be expressed 
in terms of~\eqref{eqn:treesplitQq} using the KK relation \eqref{eq:KKrelation} re-written
with the quark labels,
\begin{multline}
 \Sp^{(0)}\left(-P^+; 1_{\bar{q}}^{h_1},2^{h_2},3_q^{h_3}\right)  =  \\
  - \Sp^{(0)}\left(-P^+;1_{\bar{q}}^{h_1},3_q^{h_3},2^{h_2}\right) 
    + \Sp^{(0)}\left(-P^+;3_{\bar{q}}^{h_3},1_q^{h_1},2^{h_2}\right)
\end{multline}
A sample of two representative tree-level splitting amplitudes  $g \to \bar{q}gq$ is
\begin{align}
\Sp^{(0)}\left(-P^+;1_{\bar{q}}^+,2^-,3_{q}^-\right) &= \frac{[1P]^2}{[12] [23] s_{123}}\\[0.2in]
%
%
\Sp^{(0)}\left(-P^+;1_{\bar{q}}^-,2^+,3_{q}^+\right) &=  -\frac{\la z_1\ra ^2}{\la 12\ra  \la 23\ra } \label{eq:treesplitsub2}
\end{align}
The non-zero independent one-loop splitting amplitudes  $g \to \bar{q}qg$ are 
\begin{align}
%
%
\Sp&^{[\cN=4]} \left(-P^+;1_{\bar{q}}^+,2_{q}^-,3^-\right) = 
      \Sp^{(0)}  \left(-P^+;1_{\bar{q}}^+,2_{q}^-,3^-\right)  V_g \ \notag \\&
     -\frac{ \la z_2\ra  \la z_3\ra }{[23] \la 12\ra} \Biggl(\frac{ \left(1-z_1\right)^2 s_{12}}{ \spX_{21} z_3 s_{1P}}\Fnmhv{2} 
      -\left(\frac{\spX_{32}^2 s_{2P}^2 \left(\spX_{12}^{2}\right)^\dagger}{\left(1-z_3\right) \spX_{23} s_{12} s_{3P}} 
      -\frac{\spX_{23}}{\left(1-z_3\right) z_3}\right)\Fnmhv{1}   \notag \\&
      -\frac{z_3 s_{12} s_{13}^3}{\spX_{21} \spX_{23} \spX_{31}^2 s_{123} s_{1P}^2 s_{3P} (\spX_{31})^{\dagger }}\Biggl(\frac{z_1 \sptr_{12} s_{23}^3}{s_{12} s_{13} s_{3P} (\spX_{13})^{\dagger }}  
      -\frac{\spX_{31}^2 s_{1P} (\spX_{31})^{\dagger } s_{23}^2}{z_3 s_{13}^3} \notag \\&
      +\frac{z_1 \sptr_{32} s_{1P}}{s_{13} (\spX_{13})^{\dagger }} 
      +\frac{\spX_{23} z_3 s_{3P}}{\spX_{21} s_{13} (\spX_{31})^{\dagger }}\Biggr)\Fnmhv{3}\Biggl) \\[0.2in]
%
\Sp&^{[\cN=4]}\left(-P^+;1_{\bar{q}}^-,2_{q}^+,3^-\right) = 
      \Sp^{(0)}\left(-P^+;1_{\bar{q}}^-,2_{q}^+,3^-\right) V_g \ \notag \\ &
      -\frac{\la z_3\ra  [2P]}{[13]}\Biggl(
      -\frac{\spX_{32} z_2    s_{3P} (\spX_{23})^{\dagger } (\spX_{31})^{\dagger } \spX_{13}^2}{\spX_{23} z_3^2   s_{13} s_{23}} \Fnmhv{2} \notag \\ &
      +\frac{\spX_{32}}{\left(1-z_3\right)  \spX_{23} s_{3P}}\left(\frac{\spX_{31}  z_2^2 }{\spX_{21} }
      -\frac{\spX_{13} \spX_{21} z_1 s_{13} }{ \spX_{31}  z_3 s_{12}  (\spX_{13})^{\dagger }}\right)\Fnmhv{1}\notag \\ &
      -\left(\frac{\spX_{12} s_{13} (\spX_{31})^{\dagger } }{z_3 s_{12} s_{3P}}
      +\frac{s_{13} (\spX_{31})^{\dagger } }{z_3 s_{123} s_{3P}}
      +\frac{z_1 s_{13} }{\spX_{12} \spX_{32} s_{123} s_{3P} (\spX_{13})^{\dagger }}\right)\Fnmhv{3}\Biggl)\\[0.2in]
%
\Sp&^{[\cN=4]}\left(-P^+;1_{\bar{q}}^+,2_{q}^-,3^+\right) =\Sp^{(0)}\left(-P^+;1_{\bar{q}}^+,2_{q}^-,3^+\right) \left( V_g + F^{\text{MHV}} \right)\\[0.2in]
%
\Sp&^{[\cN=4]}\left(-P^+;1_{\bar{q}}^-,2_{q}^+,3^+\right) = \Sp^{(0)}\left(-P^+;1_{\bar{q}}^-,2_{q}^+,3^+\right) \left( V_g + F^{\text{MHV}} \right) 
\end{align}
%
%
\begin{align}
%
\Sp&^{[\rm R]}\left(-P^+;1_{\bar{q}}^+,2_{q}^-,3^-\right) =
      \Sp^{(0)}\left(-P^+;1_{\bar{q}}^+,2_{q}^-,3^-\right) \Biggl(
      -\frac{1}{\epsilon ^2}\left(\frac{\mu _R}{-s_{12}}\right)^{\epsilon }-\frac{3}{2 \epsilon } \left(\frac{\mu _R}{-s_{12}}\right)^{\epsilon } \notag \\ &
      -\frac{7}{2}-\frac{\delta_{\rm R}}{2} \Biggr)
      -\frac{\spr_{21} \la z_2\ra  \la z_3\ra }{[23] \la 12\ra } \Biggl(
      -\frac{ s_{12}}{2 \spX_{21} \spX_{31} \spr_{21} s_{123}} \Fonem
      -\frac{1}{2}  s_{123} s_{23} \hat{L}_2\left(s_{12},s_{123}\right)\notag \\ &
      +\frac{1}{2}  \left(-3 s_{123}-\frac{2 s_{23}}{\spX_{32}}\right) \hat{L}_1\left(s_{12},s_{123}\right)
      +\frac{1}{2}+\frac{s_{23}}{4 s_{12}} 
\Biggl) \\[0.2in]
%
\Sp&^{[\rm R]}\left(-P^+;1_{\bar{q}}^-,2_{q}^+,3^-\right) =
      \Sp^{(0)}\left(-P^+;1_{\bar{q}}^-,2_{q}^+,3^-\right) \Biggl(
      -\frac{1}{\epsilon ^2}\left(\frac{\mu _R}{-s_{12}}\right)^{\epsilon }-\frac{3}{2 \epsilon } \left(\frac{\mu _R}{-s_{12}}\right)^{\epsilon } \notag \\ &
      -\frac{7}{2}-\frac{\delta_{\rm R}}{2} \Biggr)
      +\frac{\la z_3\ra [2P]}{[13]s_{13}} \Biggl(
      \frac{ \sptr_{12} \sptr_{21} s_{13}}{2 \spX_{31} s_{123}} \Fonem
      -\frac{\sptr_{12} \sptr_{21} s_{13}^2 s_{23} }{2 \spX_{31} s_{123}}\hat{L}_2\left(s_{12},s_{123}\right)\notag \\ & 
      -\frac{3 \sptr_{12} \sptr_{21}  s_{13}^2 }{2 \spX_{31} s_{123}}\hat{L}_1\left(s_{12},s_{123}\right) 
      +\frac{ s_{12} s_{23} \spX_{12}^2}{2 \spX_{32} }\hat{L}_2\left(s_{23},s_{123}\right) \notag \\ &
      +\frac{ s_{12} \spX_{12}}{\spX_{32} }\left(\frac{\spX_{12} \left(2 s_{12} s_{123}+3 s_{23} s_{123}-3 s_{12} s_{23}\right)}{2 s_{12} s_{123}}+1\right) \hat{L}_1\left(s_{23},s_{123}\right) \notag \\ &
      +\frac{ 1}{2 } \left(3 \spX_{12}-\frac{s_{23}}{\spX_{32} s_{123}}-\frac{3 \left(\spX_{12}-1\right) s_{12}}{s_{123}}-1\right)\hat{L}_0\left(s_{23},s_{123}\right) \notag \\ &
      -\frac{\spX_{12}^2 s_{23} }{4 \spX_{32} s_{12}}+\frac{\spX_{12} s_{23} }{2 \spX_{32} s_{123}}+\spX_{12}+\frac{\left(1-\spX_{12}\right)s_{12}}{4  s_{123}}  -\frac{\left(s_{123}-s_{23}\right) s_{12}}{4 \spX_{32} s_{123}^2}  
 \Biggl) \\[0.2in]
%
%
%
\Sp&^{[\rm R]}\left(-P^+;1_{\bar{q}}^+,2_{q}^-,3^+\right) =
      \Sp^{(0)}\left(-P^+;1_{\bar{q}}^+,2_{q}^-,3^+\right) \Biggl(
      -\frac{1}{\epsilon ^2}\left(\frac{\mu _R}{-s_{12}}\right)^{\epsilon }-\frac{3}{2 \epsilon } \left(\frac{\mu _R}{-s_{12}}\right)^{\epsilon } \notag \\ &
      -\frac{7}{2}-\frac{\delta_{\rm R}}{2} 
      -\frac{\spr_{12}^3}{ \spr_{13}^3}\Biggl(
      -\frac{\Fonem}{2}
      -\frac{1}{2}  \left(6  s_{12}^2+s_{23}^2-6 s_{12} s_{123} \right) \hat{L}_2\left(s_{12},s_{123}\right) \notag \\ &
      +\left(3 s_{123} -2  s_{23} \right) \hat{L}_1\left(s_{12},s_{123}\right)
      +\frac{3}{2} \hat{L}_0\left(s_{12},s_{123}\right) \notag \\ &
      +\frac{ \left(s_{123}^2-s_{12}^2-s_{23}^2\right)}{2 \spr_{12}^2} \hat{L}_2\left(s_{23},s_{123}\right) 
      +\left(\frac{ s_{23}}{\spr_{12}^2} -\frac{2 s_{13}}{\spr_{12}}\right) \hat{L}_1\left(s_{23},s_{123}\right)\notag \\ &
      -\frac{\left(s_{123}+s_{23}\right)^2-s_{12}^2-2 s_{23} s_{12}}{4 \spr_{12}^2 s_{123} s_{23}}
      -\frac{6 s_{12}^2-s_{23}^2-2 s_{23} s_{12}}{4 s_{123} s_{12}}
      +\frac{s_{13}}{\spr_{12} s_{123}} 
      -\frac{3}{2}
     \Biggr)\Biggl)\\[0.2in]
%
\Sp&^{[\rm R]}\left(-P^+;1_{\bar{q}}^-,2_{q}^+,3^+\right) =
      \Sp^{(0)}\left(-P^+;1_{\bar{q}}^-,2_{q}^+,3^+\right) \Biggl(
      -\frac{1}{\epsilon ^2}\left(\frac{\mu _R}{-s_{12}}\right)^{\epsilon }-\frac{3}{2 \epsilon } \left(\frac{\mu _R}{-s_{12}}\right)^{\epsilon } \notag \\ &
      -\frac{7}{2}-\frac{\delta_{\rm R}}{2} 
      +\frac{\spr_{12}}{ \spr_{13}}\Biggl(\frac{\Fonem}{2}
      +\frac{\spr_{12}^2 s_{23}^2}{2}\hat{L}_2\left(s_{12},s_{123}\right)
      -2 \spr_{12}{}  s_{23}\hat{L}_1\left(s_{12},s_{123}\right)\notag \\ &
      +\frac{3 }{2 }  \hat{L}_0\left(s_{12},s_{123}\right)
      +\frac{\left(s_{123}-s_{12}\right)}{2 s_{123}}
      +\frac{\spr_{12}{} s_{23}}{ s_{123}}
      -\frac{\spr_{12}^2 s_{23}^2}{4 s_{12} s_{123} }\Biggr)\Biggl) 
\end{align}

\begin{align}
%
\Sp&^{[\rm f]}\left(-P^+;1_{\bar{q}}^+,2_{q}^-,3^-\right) =
      \Sp^{(0)}\left(-P^+;1_{\bar{q}}^+,2_{q}^-,3^-\right) \Biggl(
      -\frac{2}{3 \epsilon } \left(\frac{\mu _R}{-s_{12}}\right)^{\epsilon } 
      -\frac{10}{9} \Biggr)  \notag\\ &
      -\frac{\la z_2\ra  \la z_3\ra  }{3 [12] \la 23\ra }\Biggl(
      -\frac{2  s_{23}}{\spX_{23}} \hat{L}_3\left(s_{12},s_{123}\right)
      -\frac{1}{\spX_{21}}\hat{L}_2\left(s_{12},s_{123}\right)\notag\\ &
      -\frac{2 }{\spX_{21} \spX_{31} s_{123}} \hat{L}_1\left(s_{12},s_{123}\right)
      -\frac{2 }{\spX_{21} \spX_{31} s_{123} s_{23}} \hat{L}_0\left(s_{12},s_{123}\right)
      +\frac{1}{2 \spX_{21} s_{12} s_{123}}\Biggl) \\[0.2in]
%
%
%
\Sp&^{[\rm f]}\left(-P^+;1_{\bar{q}}^-,2_{q}^+,3^-\right) =
      \Sp^{(0)}\left(-P^+;1_{\bar{q}}^-,2_{q}^+,3^-\right) \Biggl(
      -\frac{2}{3 \epsilon } \left(\frac{\mu _R}{-s_{12}}\right)^{\epsilon } 
      -\frac{10}{9} \Biggr)  \notag\\ &
      +\frac{\la z_3\ra  s_{13} [2P]}{3 [13]   }\Biggl(
      2 \spX_{32}  s_{123}  \hat{L}_3\left(s_{12},s_{123}\right)
      -  \hat{L}_2\left(s_{12},s_{123}\right)\notag \\ &
      +\left(\frac{2   \spX_{12}^2}{\spX_{32} s_{123}}   
      +\frac{2  s_{123} \spX_{12}}{s_{12}^2}
      -\frac{2}{s_{12}} \right)\hat{L}_1\left(s_{12},s_{123}\right)\notag \\ &
      +\left( \frac{2 \spX_{12} \left(s_{12}+s_{123}\right)}{s_{12}^2 s_{123}}-\frac{2}{s_{12} s_{123}}\right)\hat{L}_0\left(s_{12},s_{123}\right) 
      -\frac{1}{2 s_{12}s_{123}}\Biggl) \\[0.2in]
%
\Sp&^{[\rm f]}\left(-P^+;1_{\bar{q}}^+,2_{q}^-,3^+\right) =
      \Sp^{(0)}\left(-P^+;1_{\bar{q}}^+,2_{q}^-,3^+\right) \Biggl(
      -\frac{2}{3 \epsilon } \left(\frac{\mu _R}{-s_{12}}\right)^{\epsilon } 
      -\frac{10}{9}   \notag\\ &
     -\frac{\spr_{12}^3}{3 \spr_{13}^3} \Biggl(
      \left(s_{13}^3-s_{23}^3\right) \hat{L}_3\left(s_{12},s_{123}\right) 
      -\frac{3  \left(s_{13}^2+2 s_{23} s_{13}\right)}{\spr_{12}} \hat{L}_2\left(s_{12},s_{123}\right)  \notag\\ &
      +\frac{3  s_{13}}{\spr_{12}^2} \hat{L}_1\left(s_{12},s_{123}\right) 
      -\hat{L}_0\left(s_{12},s_{123}\right)
      -\frac{1}{2} \frac{\left(s_{123}-s_{13}\right)^2-s_{13}s_{123}-s_{12}(s_{123}-s_{23})}{s_{12} s_{123}} \notag\\ &
      -\frac{s_{13}}{\spr_{12}^2 s_{123}}
      +\frac{s_{13} \left(3 s_{12}+5 s_{123}+s_{23}\right)}{2 \spr_{12} s_{12} s_{123}}
\Biggr)
 \Biggl) \\[0.2in]
%
\Sp&^{[\rm f]}\left(-P^+;1_{\bar{q}}^-,2_{q}^+,3^+\right) =
      \Sp^{(0)}\left(-P^+;1_{\bar{q}}^-,2_{q}^+,3^+\right) \Biggl(
      -\frac{2}{3 \epsilon } \left(\frac{\mu _R}{-s_{12}}\right)^{\epsilon } 
      -\frac{10}{9}   \notag\\ &
      -\frac{\spr_{12} }{3 \spr_{13}}\Biggl(
      -2 \spr_{12}^2s_{23}^3 \hat{L}_3\left(s_{12},s_{123}\right) 
      +3 \spr_{12}  s_{23}^2 \hat{L}_2\left(s_{12},s_{123}\right)
      +2  s_{13}\hat{L}_1\left(s_{12},s_{123}\right) \notag\\ &
      -\frac{ s_{23}}{s_{12}-s_{123}} \hat{L}_0\left(s_{12},s_{123}\right)
      +\frac{s_{23}}{s_{123}}
      -\frac{\spr_{12} s_{23}^2}{2 s_{12} s_{123}}
\Biggr)
 \Biggl) 
\end{align}

\begin{align}
%
\Sp&^{[\text{scalar}]}\left(-P^+;1_{\bar{q}}^+,2_{q}^-,3^-\right) = 
      \Sp^{(0)}\left(-P^+;1_{\bar{q}}^+,2_{q}^-,3^-\right) \left(\frac{2}{3} \delta_{\rm R}\right)  \notag\\ &
     +\frac{3 \la z_2\ra  \la z_3\ra   }{\spX_{21} \spX_{32} [12] \la 23\ra  s_{123}}\hat{L}_1\left(s_{12},s_{123}\right)
       \\[0.2in]
%
%
%
\Sp&^{[\text{scalar}]}\left(-P^+;1_{\bar{q}}^-,2_{q}^+,3^-\right) =
      \Sp^{(0)}\left(-P^+;1_{\bar{q}}^-,2_{q}^+,3^-\right) \left(\frac{2}{3} \delta_{\rm R} -2 \right) \notag\\ &
 \frac{\la z_3\ra }{[13] \la 2P\ra } \Biggl(
+\frac{3  s_{2P}}{2 \spX_{31} s_{123}}\Fonem
-\frac{3  s_{13} s_{2P}}{\spX_{31} s_{123}} \hat{L}_1\left(s_{12},s_{123}\right)
-\frac{3  s_{13} s_{2P}}{\spX_{31} s_{123}} \hat{L}_1\left(s_{23},s_{123}\right) \notag\\ &
+\frac{2 s_{13} s_{2P}}{\spX_{32} s_{12} s_{123}}  
-\frac{2 \spX_{13} \spX_{32} z_2 s_{2P}}{\spX_{23} \left(1-z_3\right) s_{12}}
+\frac{2 \spX_{12} \spX_{13} z_2 s_{2P}}{\spX_{23} z_3 s_{12}}
 \Biggr) \\[0.2in]
%
%
\Sp&^{[\text{scalar}]}\left(-P^+;1_{\bar{q}}^+,2_{q}^-,3^+\right) =
      \Sp^{(0)}\left(-P^+;1_{\bar{q}}^+,2_{q}^-,3^+\right) \Biggl(
      \frac{2}{3} \delta_{\rm R} 
      -\frac{3\spr_{12} }{\spr_{13}^2}\Biggl(
	\frac{1}{2}\Fonem   \notag\\ &
      + s_{12} \hat{L}_1\left(s_{23},s_{123}\right) 
      + s_{23} \hat{L}_1\left(s_{12},s_{123}\right)
      + \hat{L}_0\left(s_{12},s_{123}\right)
      + \hat{L}_0\left(s_{23},s_{123}\right)\Biggr)      
 \Biggl)  \\[0.2in]
%
%
\Sp&^{[\text{scalar}]}\left(-P^+;1_{\bar{q}}^-,2_{q}^+,3^+\right) =
      \Sp^{(0)}\left(-P^+;1_{\bar{q}}^-,2_{q}^+,3^+\right) \Biggl(
      \frac{2}{3} \delta_{\rm R} -3 \spr_{12} s_{23} \hat{L}_1\left(s_{12},s_{123}\right)
 \Biggl) 
\end{align}

The expressions for the non-zero independent one loop splitting amplitudes $g \to \qb g q$ are
\begin{align}
%
%
\Sp&^{[\cN=4]} \left(-P^+;1_{\bar{q}}^+,2^-,3_{q}^-\right) = 
      \Sp^{(0)}  \left(-P^+;1_{\bar{q}}^+,2^-,3_{q}^-\right)\Biggl(  V_g 
      +\frac{\spX_{21} \left(1-z_1\right)^2  s_{123}}{z_2 s_{1P}}\Fnmhv{2} \notag \\&
      +\left(\frac{\spX_{21} \spr_{23} s_{23} (\spX_{13})^{\dagger }}{z_1 s_{1P}}+1\right)\Fnmhv{3}
      -\frac{\spX_{31}^2  s_{123} \left(\spX_{13} \spr_{23}\right)^{\dagger }}{s_{13}} \Fnmhv{1}     
     \Biggl) \\[0.2in]
%
%
\Sp&^{[\cN=4]}\left(-P^+;1_{\bar{q}}^-,2^+,3_{q}^+\right) = \Sp^{(0)}\left(-P^+;1_{\bar{q}}^-,2^+,3_{q}^+\right) \left( V_g + F^{\text{MHV}} \right) 
\end{align}

\begin{align}
%
%
\Sp&^{[\text{scalar}]}\left(-P^+;1_{\bar{q}}^+,2^-,3_{q}^-\right) =
      \Sp^{(0)}\left(-P^+;1_{\bar{q}}^+,2^-,3_{q}^-\right) \Biggl(
      -\frac{2}{3 \epsilon } \left(\frac{\mu _R}{-s_{13}}\right)^{\epsilon } 
      -\frac{10}{9} \notag\\ &
      + \frac{2 \delta_{\rm R}}{3} 
      +\frac{3 \spX_{21}  s_{23}}{\spX_{23}} \hat{L}_1\left(s_{12},s_{123}\right)
      -\frac{2}{3} \hat{L}_0\left(s_{13},s_{123}\right)
 \Biggl) \\[0.2in]
%
\Sp&^{[\text{scalar}]}\left(-P^+;1_{\bar{q}}^-,2^+,3_{q}^+\right) =
      \Sp^{(0)}\left(-P^+;1_{\bar{q}}^-,2^+,3_{q}^+\right) \Biggl(
      -\frac{2}{3 \epsilon } \left(\frac{\mu _R}{-s_{123}}\right)^{\epsilon } 
      -\frac{29}{18} \notag\\ &
      + \frac{2 \delta_{\rm R}}{3} 
      + \left(\frac{\spX_{32}^2 s_{12}^2}{2 \spX_{12}^2}-\frac{s_{13}^2}{2 \spX_{13}^2}\right)\hat{L}_2\left(s_{12},s_{123}\right)
      -4 \spr_{12}  s_{23} \hat{L}_1\left(s_{12},s_{123}\right) \notag\\ &
      +\frac{1}{2} \hat{L}_0\left(s_{12},s_{123}\right)
      +\frac{ \spX_{32}^2 s_{12}}{4 \spX_{12}^2 s_{123}}
      -\frac{\spX_{32}}{\spX_{12}}
      -\frac{s_{13}^2}{4 \spX_{13}^2 s_{12} s_{123}}
      +\frac{s_{123}}{2 s_{12}}
 \Biggl)
\end{align}
The expressions for the others helicity configurations are obtained by conjugation operation or by reverting the fermion line, namely 
$\Sp\left(-P;1_{\bar{q}}^{h},2,3_{q}^{-h}\right) = \Sp\left(-P;3_{\bar{q}}^{-h},2,1_{q}^{h}\right)$. 

\section{Super-symmetric Ward identities}

Super-symmetric Ward identities (SWI) allow us to relate amplitudes with different particle content in
super-symmetric theories \cite{Mangano:1987xk,Mangano:1990by}. We have checked that the super-symmetric primitive amplitudes
presented in the previous sections obey the expected SWI which are easy to derive by taking the
triple collinear limit of the equations presented \cite{Mangano:1990by}. The resulting equations
differ for the MHV and NMHV helicity configurations,
\begin{align}
  \Sp^{[\cN=4]}(-P^+,1^+,2^{-},3^+)
  &- \frac{\zspA2}{\zspA1}\Sp^{[\cN=4]}(-P^+,1_{\bar{q}}^+,2_q^{-},3^+) = 0, \\
   \Sp^{[\cN=4]}(-P^+,1^-,2^{+},3^+)
  &+ \frac{\zspA1}{\zspA2}\Sp^{[\cN=4]}(-P^+,1_{\bar{q}}^-,2_q^{+},3^+) = 0, \\
  \Sp^{[\cN=4]}(-P^+,1^+,2^-,3^-)
  &- \frac{\zspA2}{\zspA1}\Sp^{[\cN=4]}(-P^+,1_{\bar{q}}^+,2_q^-,3^-) \nonumber\\&
   - \frac{\zspA3}{\zspA1}\Sp^{[\cN=4]}(-P^+,1_{\bar{q}}^+,2^-,3_q^-)
  = 0, \\
  \Sp^{[\cN=4]}(-P^+,1^-,2^+,3^-)
  &+ \frac{\zspA1}{\zspA2}\Sp^{[\cN=4]}(-P^+,1_{\bar{q}}^-,2_q^+,3^-) \nonumber\\&
   + \frac{\zspA3}{\zspA2}\Sp^{[\cN=4]}(-P^+,3_{\bar{q}}^-,2_q^+,1^-)
  = 0.
\end{align}
The SWI are also applicable to tree-level amplitudes in non-super-symmetric theories.

\section{Cross checks}
\label{sec:crosschecks}

We check the universality of the splitting functions derived in the
previous section numerically against the six parton
amplitudes available in \textsc{NJet} \cite{Badger:2012pg}. In order
to make sure we could evaluate as close to the precise limit as
possible, we implemented the checks in octuple precision using the
\textsc{qd} and \textsc{OneLoop} \cite{vanHameren:2010cp} packages.

We check the validity of Eq.~\eqref{eq:csumfactlimit} by computing the
ratio between the two sides of the equation summed over the external
helicities
\begin{align}
  & r_{\mathrm{collinear } 123} = \nonumber \\
& \quad   \frac{\sum_{\hh_i} \mathcal{M}_n^{(L,L')}(\{p_i^{\hh_i}\}) }{
  \sum_{k=0}^L \sum_{k'=0}^{L'} \sum_{\hh_P, \hh_P'}
  \mathcal{P}^{(L-k,L'-k')}_{m,s;-\hh_P,-\hh_P'}(-P; \{p_i\}_{i=1}^m)
  \, \mathcal{M}_{n-m+1,s;\hh_P,\hh_P'}^{(k,k')}(\{P, \{p_i\}_{i=m+1}^n\}) } \label{eq:colorsumratio}
\end{align}
where $\mathcal{M}_{n,s}$ and $\mathcal{P}_{n,s}$ in the denominator
are defined from $\mathcal{M}_{n}$ and $\mathcal{P}_{n}$ by summing
over the external helicities:
\begin{align}
\mathcal{M}^{(L,L')}_{n-m+1;\hh_P,\hh_P'}(P, \{p_i\}) ={}& \sum_{\hh_{m+1},\ldots ,\hh_{n}}\mathcal{M}^{(L,L')}_{n-m+1,s;\hh_P,\hh_P'}(P, \{p_i^{\hh_i}\}) \nonumber \\
\mathcal{P}_{m;\hh_P,\hh_P'}^{(L,L')}(-P; \{p_i\}) ={}& \sum_{\hh_{1},\ldots ,\hh_{m}}\mathcal{P}_{m,s;\hh_P,\hh_P'}^{(L,L')}(-P; \{p_i^{\hh_i}\}) \nonumber .
\end{align}
Eq.~\eqref{eq:csumfactlimit} obviously implies
\begin{equation}
  r_{\mathrm{collinear } 123} \xrightarrow{1||2||3} 1. \label{eq:colorsumratioto1}
\end{equation}
It is worth observing that the finite one-loop all-plus and all-minus
four-gluon helicity amplitudes, while giving no contribution to the
NLO squared matrix element, they give instead a finite contribution to
$r_{\mathrm{collinear } 123}$ because of spin correlations.

\begin{figure}[t]
  \centering
  \includegraphics[width=0.49\textwidth]{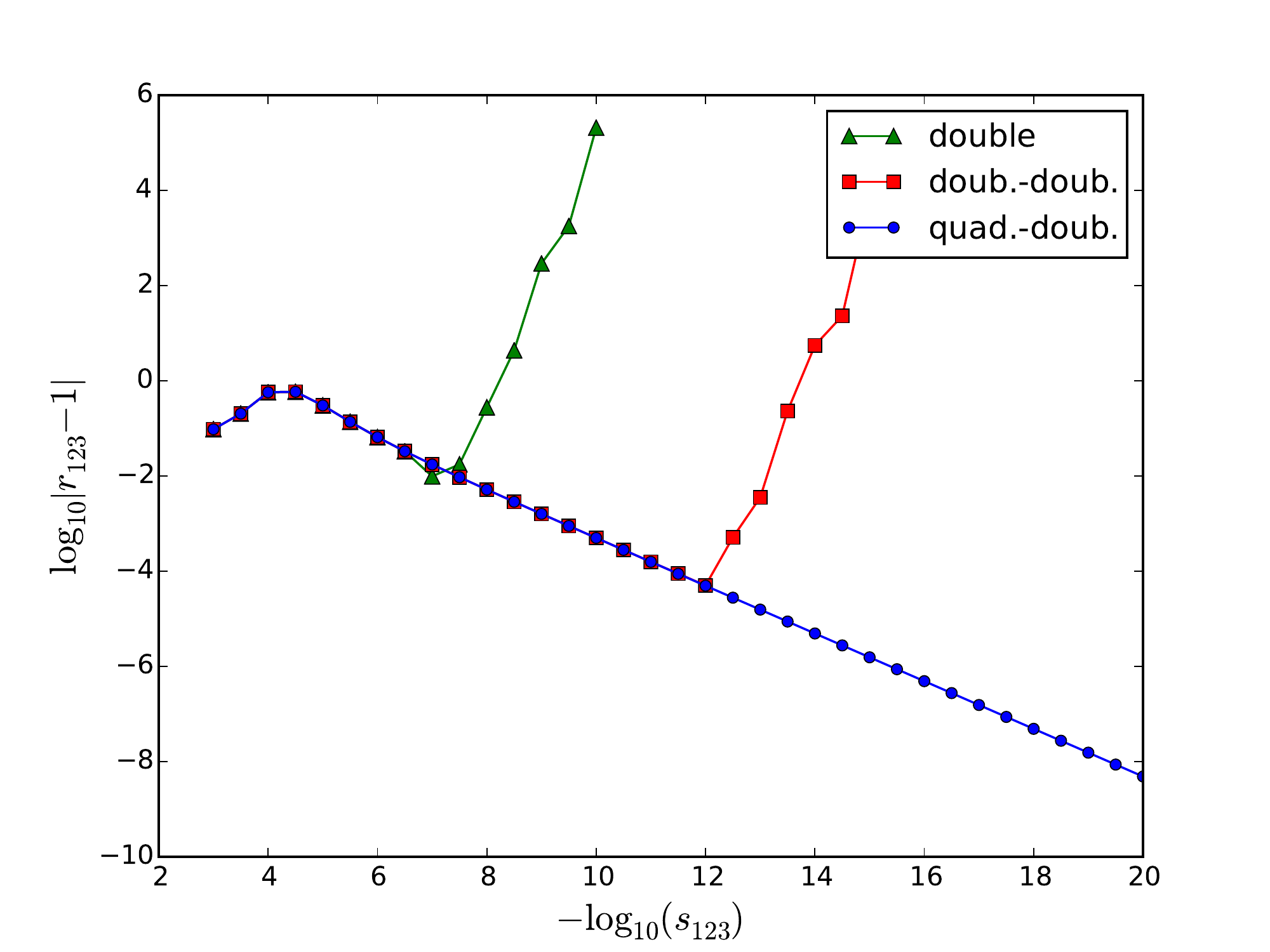}
  \includegraphics[width=0.49\textwidth]{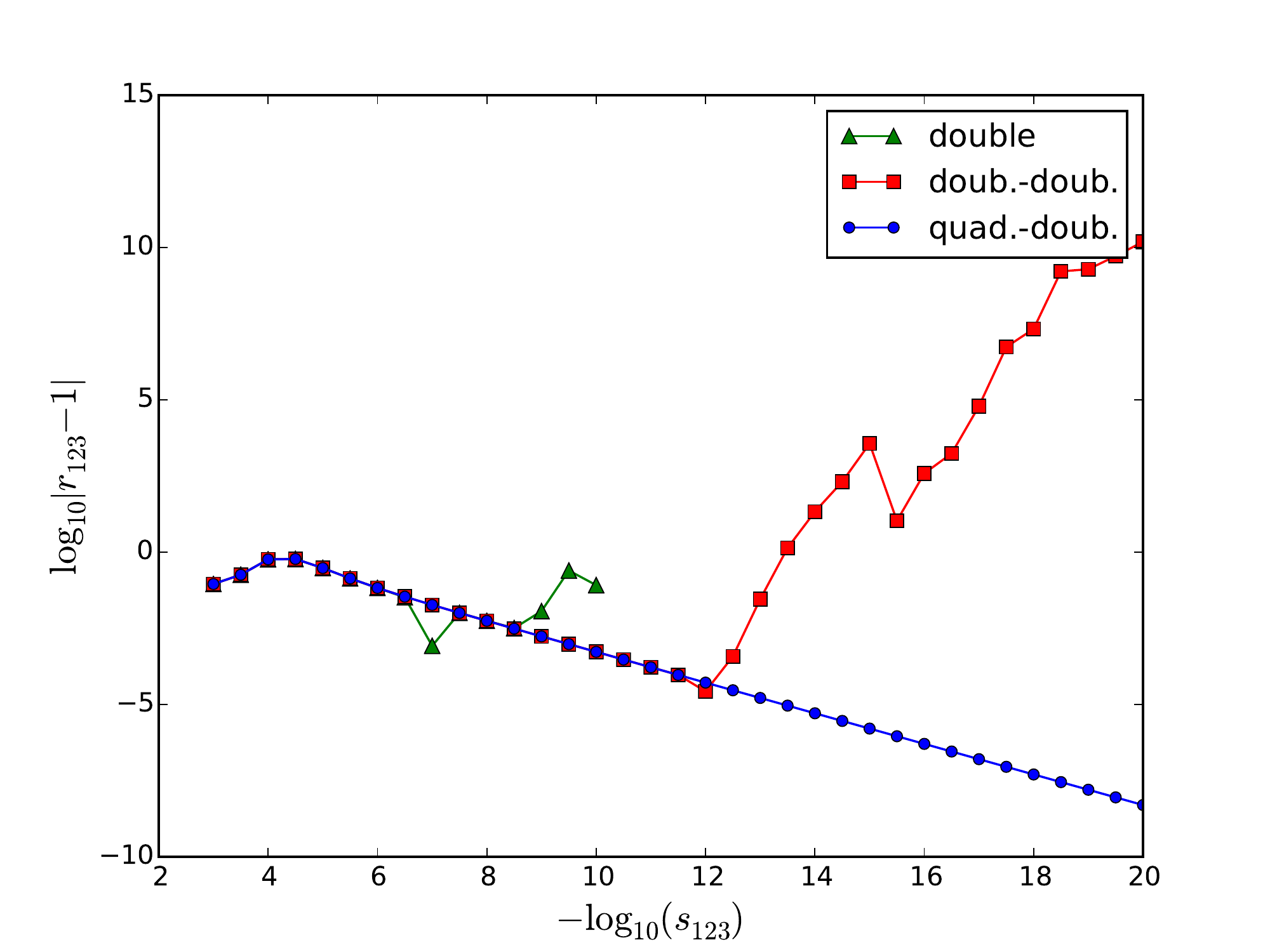}
  \caption{Numerical check of the collinear limit of
    $r_{\mathrm{collinear } 123}-1$, with
    $r_{\mathrm{collinear } 123}$ defined by
    Eq.~\eqref{eq:colorsumratio}, as a function of the invariant mass
    $s_{123}$, from $s_{123}=10^{-3}$ to $s_{123}=10^{-20}$.  For this
    check we set $n_f=5$, $\sqrt{s}=10^3$ and $\mu_R=10^3/7$.  The
    plot on the left shows the all-gluon case, while the one on the right
    shows the quark case.  In the latter it was not possible to obtain
    numerical results in double precision for
    $s_{123}\lesssim 10^{-10}$.}
  \label{fig:colorsum}
\end{figure}

In Fig.~\ref{fig:colorsum} we plot $r_{\mathrm{collinear } 123}-1$ as
a function of the invariant mass $s_{123}$ of the three collinear
partons.  More in detail we verify the validity of
Eq.~\eqref{eq:colorsumratioto1} in double, double-double and
double-quadruple precision for both gluon (on the left) and quark (on
the right) splitting functions.  As one can see, going to higher
precision allowed us to make stronger checks on phase-space space
points which are closer to the limit, where the numerical evaluation
is highly unstable at lower floating-point precision.

Similarly, we also numerically verified Eq.~\eqref{eq:collfact} for
each primitive amplitude and all the helicity configurations, although
all of these already contribute to the check described above.

As well as the numerical checks we have also verified that all splitting functions factorise
correctly in the iterated collinear limit,
\begin{align}
  \underset{1||2}{\lim} \,\Sp^{(L)}&(-P_{123}^{-\lambda_P}, 1^{\lambda_1}, 2^{\lambda_2}, 3^{\lambda_3})
  \nonumber\\ &
  = \sum_{\lambda=\pm} \sum_{k=0}^{L}
  \Sp^{(L-k)}(-P_{\widetilde{12}3}^{-\lambda_P}, \tilde{P}_{12}^{\lambda}, 3^{\lambda_3})
   \Sp^{(k)}(-P_{12}^{-\lambda}, 1^{\lambda_1}, 2^{\lambda_2}),
  \label{eq:iteratedlimit}
\end{align}
where the scale $s_{12} \ll s_{123}$ and $P_{\widetilde{12}3}=\tilde{P}_{12}+p_3$. All di-logarithms drop out in this limit though some
care should be taken to ensure the hierarchy of scales is imposed correctly.

\section{Conclusions}

In this article we have computed the one-loop triple collinear splitting amplitudes in QCD
initiated by a gluon. These functions are one of the last remaining ingredients to complete
the classification of universal infrared limits relevant at N${}^3$LO.

Some effort has been taken to ensure the splitting amplitudes have compact analytic forms.
We made use of the spinor-helicity formalism and super-symmetric decompositions and related the
pure gluonic amplitudes to the ones containing a quark anti-quark pair. The primitive amplitude colour decomposition was also
a useful tool to express full colour and helicity summed splitting functions which were all checked
explicitly against the numerical matrix elements for $2\to4$ scattering in \textsc{NJet}. In the
course of these checks we made use of the high precision numerical evaluation available with
up to 64 digits via the \textsc{qd} package. This allowed us to probe deep into the collinear limit
and verify that all parts of the computation behaved correctly. This was particularly important
for the spin correlated and sub-leading colour corrections which are significantly suppressed.

There are still some missing ingredients needed for the constructions of a fully differential N${}^3$LO subtraction
scheme. Firstly, the quark initiated channels are still unavailable - they are not directly
accessible from the $H+2j$ amplitudes since they have been computed in the effective theory where
the Higgs couples only to gluons. The necessary splitting amplitudes could be extracted from the
vector boson plus four parton one-loop amplitudes \cite{Bern:1997sc}.

Secondly when integrating the splitting functions over the unresolved phase space the expansion of
the limit may be required to higher order in the dimensional regularisation parameter $\eps$. This
would require a new computation of the one-loop matrix elements valid in $D=4-2\eps$ dimensions which is
quite feasible using modern unitarity methods. The appearance of the one-loop pentagon function in
the full $D$-dimensional amplitude may complicate this part of the computation even if it is only
required in the triple collinear limit.

We hope that the expressions presented here will be of use in future high precision QCD
computations.

\begin{acknowledgments}
  We are grateful to Franz Herzog, Tom Melia and Einan Gardi for useful discussions. The work of
  S.B. is supported an STFC Rutherford Fellowship ST/L004925/1.
\end{acknowledgments}

\appendix

\section{Generation of collinear phase space points \label{app:collinearphasespace}}

In this Appedix we illustrate a practical way to generate a set of
on-shell $n$-particle phase-space points where the first $m$ particles
approach the collinear limit $1||\cdots|| m $.  The limit is
approached by varying a single free parameter $\delta$ as
$\delta \to 0$ and it is based on the parametrisation presented in
Section~\ref{sec:limparam}.  This has been used for the numerical
checks we discussed in Section~\ref{sec:crosschecks}.

As a first step we generate an on-shell $(n-m+1)$-particle phase space
point defining the set of momenta
\begin{equation}
  \{\tP,\, p_{m+1}(0),\, p_{m+2}(0), \ldots,\, p_n(0)\}
\end{equation}
where, as suggested by the notation, $p_i(0)$ for $i\geq m+1$ are the
momenta of the non-collinear particles at $\smallpar=0$, while $\tP$ is
the sum of the collinear momenta in the limit.  We then define the
\emph{exact} collinear limit as the set of momenta
\begin{equation}
  \{ z_1 \tP,\, z_2 \tP, \ldots ,\, z_m \tP,\,   p_{m+1}(0), \ldots , \, p_n(0) \},
\end{equation}
where $z_i$ are randomly generated real numbers satisfying
Eq.~\eqref{eq:sumzi}.  In order to avoid regions with soft kinematics
(which would introduce other kinds of singularities) one can generate
a set of random numbers between, for example, 1 and 3 and divide them by their sum.

In order to define the orthogonal direction we must specify the
reference vector $\rvec$ appearing in Eq.~\eqref{eq:pidelta}.  A
particular convenient choice is one of the non-collinear vectors,
i.e.\
\begin{equation}
  \rvec^\mu = p_{m+1}^\mu(0).
\end{equation}
The orthogonal direction is thus spanned by the two complex vectors
\begin{equation}
  \frac{\spAB{\tP}{\gamma^\mu}{\rvec}}{2}, \quad \frac{\spAB{\rvec}{\gamma^\mu}{\tP}}{2}.
\end{equation}
While these are particularly convenient when working with the
spinor-helicity formalism, for numerical checks with real kinematics
it is convenient to define two real linear combinations
\begin{equation}
  v_{1,\perp}^\mu = \frac{1}{2}\, \left(\frac{\spAB{\tP}{\gamma^\mu}{\rvec}}{2} + \frac{\spAB{\rvec}{\gamma^\mu}{\tP}}{2} \right), \quad  v_{2,\perp}^\mu = \frac{1}{2\, i}\, \left( \frac{\spAB{\tP}{\gamma^\mu}{\rvec}}{2} -\frac{\spAB{\rvec}{\gamma^\mu}{\tP}}{2} \right).
\end{equation}
Hence the orthogonal vectors $k^\mu_{T,i}$ are defined as
\begin{equation}
  k^\mu_{T,i} =  y_{1,i}\, v_{1,\perp}^\mu + y_{2,i}\, v_{2,\perp}^\mu \label{kTdefpsp}
\end{equation}
where $y_{1,i}$ and $y_{2,i}$ are randomly generated real numbers
satisfying
\begin{equation}
  \sum_i y_{1,i} = \sum_i y_{2,i} = 0.
\end{equation}
The variables $y_{1,i}$ and $y_{2,i}$ are related
to the spinor variables $\zspA{i}$, $\zspB{i}$, $\wspA{i}$ and
$\wspB{i}$ introduced in Section~\ref{sec:limparam} by
\begin{equation}
  y_{1,i} = \zspA{i}\,  \wspB{i} + \wspA{i} \, \zspB{i}, \quad y_{2,i} = i\, (\zspA{i}\,  \wspB{i} - \wspA{i} \, \zspB{i}),
\end{equation}
as one can check by requiring consistency with Eq.~\eqref{eq:kTdef}.
As already stated, these spinor variables differ by a phase from the
usual parametrisation in terms of $\sqrt{z_i}$.  If
\begin{equation}
  \zspA{i} = \zspB{i}^* = \sqrt{z_i} \, e^{i\, \theta}, \qquad \wspA{i} =  \wspB{i}^* = \sqrt{\omega_i}\, e^{i\, \phi}
\end{equation}
then
\begin{equation}
  y_{1,i} = 2\, \sqrt{z_i\, \omega_i} \cos(\phi-\theta), \quad y_{2,i} = 2\, \sqrt{z_i\, \omega_i} \sin(\phi-\theta).
\end{equation}
From here it is easy to see that while the parametrisation in terms of $\zspA{i}$
and $\zspB{i}$ has the advantage of producing results that are
analytic functions of the spinor variables in the complex plane, the
parametrisation in terms of $\sqrt{z_i}$ is in fact entirely equivalent in
the physical region.

Using $k^\mu_{T,i}$ as in Eq.~\eqref{kTdefpsp}, one can simply define
the momenta $p_1,\ldots , p_m$ for any value of the free parameter
$\smallpar$ using Eq.~\eqref{eq:pidelta}.  With our choice of $\rvec$
we can absorb the recoil by defining
\begin{align}
  p_{m+1}^\mu(\smallpar) ={}& \Big(1+\sum_{i=0}^{m} \frac{\smallpar^2 \, k_{T,i}^2}{2\, z_i\, (\tP \cdot \rvec)}\Big)\, p_{m+1}^\mu (0) \\
  p_{i}^\mu(\smallpar) = {}& p_i^\mu(0), \qquad m+2 \leq i \leq n.
\end{align}

\providecommand{\href}[2]{#2}\begingroup\raggedright\endgroup
\end{document}